\documentclass[aps,pre,superscriptaddress,showpacs]{revtex4}
\usepackage{graphicx}
\usepackage{epsfig}
\usepackage{amsfonts}
\usepackage{color}
 \newcommand{\beq}{\begin{equation}}
 \newcommand{\eeq}{\end{equation}}
 \newcommand{\beqa}{\begin{eqnarray}}
 \newcommand{\eeqa}{\end{eqnarray}}

\newcommand{\be}{\begin{equation}}
\newcommand{\ee}{\end{equation}}
\newcommand{\bea}{\begin{eqnarray}}
\newcommand{\eea}{\end{eqnarray}}
\newcommand{\bay}{\begin{array}}
\newcommand{\eay}{\end{array}}
\newcommand{\bfig}{\begin{figure}}
\newcommand{\efig}{\end{figure}}
\newcommand{\bc}{\begin{center}}
\newcommand{\ec}{\end{center}}
\newcommand{\btab}{\begin{tabular}}
\newcommand{\etab}{\end{tabular}}
\newcommand{\dr}{\partial}

\newcommand{\szz}{\sigma_{zz}}
\newcommand{\sxx}{\sigma_{xx}}

\newcommand{\sxz}{\sigma_{xz}}
\newcommand{\sij}{\sigma_{ij}}

\newcommand{\uzz}{u_{zz}}
\newcommand{\uxx}{u_{xx}}

\newcommand{\uxz}{u_{xz}}

\newcommand{\iqt}{\int_{-\infty}^{+\infty} \!\!\!\!\!\!\!\!\! dq \,}
\newcommand{\iqz}{\int_{0}^{+\infty} \!\!\!\!\!\!\!\!\! dq \,}
\newcommand{\cz}{\cos\theta_0}
\newcommand{\sz}{\sin\theta_0}

\newcommand{\StartTwoColumn}{\begin{multicols}{2}}
\newcommand{\EndTwoColumn}{\end{multicols}}

 \renewcommand{\vec}{\bf }
 
\begin{document}

\title{Anisotropy in granular media: classical elasticity and directed force chain network}
\author{M. Otto}
\affiliation{Institut f\"ur Theoretische Physik, Universit\"at
  G\"ottingen, Bunsenstr. 9, D-37075 G\"ottingen, Germany}
\author{J.-P. Bouchaud}
\affiliation{Service de Physique de l'Etat Condens\'e, CEA-Saclay,
  Orme des Merisiers, 91191 Gif-sur-Yvette Cedex, France}
\author{P. Claudin}
\affiliation{Laboratoire des Milieux D\'esordonn\'es et
  H\'et\'erog\`enes (UMR 7603), 4 place Jussieu -- case 86, 75252 Paris
  Cedex 05, France}
\author{J. E. S. Socolar}
\affiliation{Department of Physics and Center for Nonlinear and
  Complex Systems, Duke University, Durham, NC 27708, USA}

\begin{abstract}
  A general approach is presented for understanding the stress
  response function in anisotropic granular layers in two dimensions.
  The formalism accommodates both classical anisotropic elasticity
  theory and linear theories of anisotropic directed force chain
  networks.  Perhaps surprisingly, two-peak response functions can
  occur even for classical, anisotropic elastic materials, such as
  triangular networks of springs with different stiffnesses.  In such
  cases, the peak widths grow linearly with the height of the
  layer, contrary to the diffusive spreading found in
  `stress-only' hyperbolic models.  In principle, directed force chain
  networks can exhibit the two-peak, diffusively spreading response
  function of hyperbolic models, but all models in a particular class
  studied here are found to be in the elliptic regime.
\end{abstract}

\maketitle
\section{Introduction}
The stress response of an assembly of hard, cohesionless grains has
been a subject of debate \cite{deG,Savage,us,Proc}. The dividing line
has been mostly between traditional approaches based on elasticity or
elastoplasticity theory on one hand, and ``stress-only'' models on the
other which make no reference to a local deformation field but posit
(history-dependent) closure relations between components of the stress
tensor.  The former leads to elliptic partial differential
equations for the stresses, for which boundary conditions must be
imposed everywhere on the boundary. In contrast, the latter
approach often leads to hyperbolic equations \cite{us,Proc}.
The wave-like behavior of their solutions has been at the origin of a
proposed physical mechanism called stress propagation through the bulk
a granular material. In an infinite slab geometry, it only requires the specification of boundary
conditions on the ``top'' surface. A family of (linear) closure
relations have been shown to account for the pressure dip underneath
the apex of a sandpile and stresses in silos \cite{us,Proc}. Alternative
explanations based on elastoplasticity are found in \cite{Cantelaube}.

The phenomenological `stress-only' closure relations follow from
plausible symmetry arguments, and can be seen as the
  coarse-grained version of local probabilistic rules
for stress transfer \cite{pre}.  
However,
these relations lack a detailed
microscopic derivation that would allow one both to understand their
range of validity and to compute the phenomenological parameters from
the statistical properties of the packing, except in the case of
frictionless grains. In fact, a system of frictionless polydisperse
spheres is shown to be isostatic \cite{bcc, Roux, moukarzel, Witten},
i.e. the number of unknown forces is equal to the number of equations
for mechanical equilibrium. If an isostatic
system is sufficiently anisotropic, a linear closure relation between
stresses can be derived \cite{Edwards}.
Further attempts to obtain the missing equation for stresses from a
microscopic approach for different packings are presented in \cite{Edwards, Ball}, but these are still
somewhat inconclusive. In particular, in the case of a completely
  isotropic packing, none of the homogeneous linear closure relations
  is compatible with the rotational symmetry. The idea of `grains' (in
  the metallurgical sense) and packing defects must be introduced to
  restore the large scale symmetry.

In order to understand stress distribution on a more fundamental
level, we have introduced the mesoscopic concept of the directed force
chain network ({\sc dfcn}) \cite{bclo,ssc}, which is
motivated by the experimental evidence for filamentary force
chains in a wide variety of systems.\cite{photo}. The ``double Y''
model has been developed to describe such networks based on simple
rules for the splitting and merging of straight force chains.
This model leads to a non-linear Boltzmann equation for the
probability $P(f,\hat{\bf n},{\bf r})$ of finding a force chain
at the spatial point ${\bf r}$ with intensity $f$ in the direction
$\hat{\bf n}$.

In a first paper \cite{bclo}, chain merging (which produces the
non-linear terms in the Boltzmann equation) was neglected. An
isotropic splitting rule was assumed, corresponding to strongly
disordered isotropic granular packings.  A pseudo-elastic theory for
the stress tensor was derived in which the role of the displacement
field is played by a vector field ${\bf J}({\bf r})=\langle \hat{\bf
  n} f \rangle$ that represents the coarse-grained or ensemble
averaged force chain direction.  A relation between $\partial_i J_j$
and the stress tensor exists that is formally equivalent to an
isotropic stress-strain relation.  The resulting elliptic
  equations yield a response function with a unique (pseudo-elastic)
  peak, as observed experimentally in strongly disordered packings
  \cite{Reydellet,Reydelletbis}.  Further study showed, however, that
  the non-linear terms in the Boltzmann equation contain essential
  physics and cannot be neglected. \cite{ssc} In fact, for an exactly
solvable model with 6 discrete directions for force chains, it was
found that the elliptic (pseudo-elastic) behavior of the response
function is limited to small depths, and that at sufficiently
large depths a crossover occurs to an hyperbolic response -- two
Gaussian peaks that propagate away and broaden diffusively.
  Whether this behavior is specific to the model with 6
discrete directions is a subject of current study, and the elliptic or
hyperbolic nature of the linearized response around the full solution
of the nonlinear Boltzmann equation is an open question.

Following a different route, Goldenberg and Goldhirsch
\cite{goldhirsch} have recently noted that a two-peak response
function can be found in classical anisotropic ball-and-spring models.
Gay and da Silveira \cite{Gay} have furthermore given some arguments
for the relevance of anisotropic elasticity for the large scale
description of granular assemblies of compressible grains that can
locally rotate.  The two-peak nature of the response function is
therefore not, in itself a signature of hyperbolicity, but
may occur in elliptic systems that are sufficiently
anisotropic.  The unambiguous signature of hyperbolic response
  lies in the scaling of the peak widths with depth, which is linear
  in generic elliptic systems but diffusive (proportional to the
  square root of depth) in generic hyperbolic systems.  In the linear
  pseudo-elasticity theories discussed below, the diffusive spreading
  in hyperbolic systems is not captured; the peaks appear as delta
  functions that do not spread at all. Deviations from elasticity on small scales and their possible relation
with
granular media were also discussed in \cite{Tanguy}.

The aim of this paper is to give a unified account of the shape
of the response function for anisotropic systems described either
  by standard elasticity theory or the pseudo-elastic theory that
  emerges from an approximate linear treatment of directed force
  networks.  Though there are open questions concerning the
  self-consistency of the latter, there do appear to be some contexts
  in which the equations of the pseudo-elasticity theory hold, and
  they may be especially relevant for systems of intermediate depth
  (large compared to the disorder length scale but not much larger
  than the persistence length of force chains).

Very recently, the response functions of two-dimensional granular
layers subjected to shear have been determined experimentally
\cite{geng}.  Under shear, an anisotropic texture appears and force
chains are preferably oriented along an angle of 45 degrees.  Within
the (pseudo)-elasticity framework presented below, this provides
motivation for studying materials characterized by a selected global
direction ${\bf N}$.

The paper is organized as follows. In section 2 a general
mathematical framework for calculating stress response
  functions in anisotropic materials.  The main results of the paper
  are then summarized in a `phase-diagram' indicating where `one-peak'
  or `two- peak' response functions can appear in parameter space.  In
  section 3, we compute the analytic form of the response function for
  the various phases and show a number of examples of the variety of
  shapes that are possible, including a brief comment on relation to
  experimental work.  In section 4, we show how the formalism
  applies to the example of a triangular ball-and-spring network,
  indicating how spring stiffnesses must be chosen to access all
  possible regions of the general parameter space.  In section 5, a
  linear anisotropic pseudo-elastic theory is derived from an
  anisotropic linear directed force chain network model and it is
  shown that this class of models always lies in the elliptic
  regime.  A conclusion is given in section 6. Algebraic
  details of several calculations are presented in Appendices.

\section{Anisotropic elasticity and summary of our results}
\label{ela.general}
\subsection{General equations for 2D systems with arbitrary\\ anisotropy}

In the following, we present a general framework that covers both
classical linear anisotropic elasticity theory \cite{LL} and a
generally anisotropic ``pseudo-elasticity'' theory, that appears
within a linearized treatment of directed force chain networks (see
section \ref{dfcn}).  The large scale equations that can be derived in
these two approaches are formally identical, although the
``pseudo-strain'' has a geometric meaning different from the usual
  strain tensor. For simplicity, we will restrict the
  discussion to two-dimensional systems.

The most general linear relation between the stress tensor ${\bf\sigma}$ and a
symmetric tensor formed from the gradients of a vector field ${\bf u}$ is
\beq
\label{s.u}
\sigma_{ij}=\lambda_{ijkl}u_{kl}, \eeq where $\sigma_{ij}$ denotes
a component of the stress tensor,
$u_{ij}\equiv\frac{1}{2}(\partial_j u_i+\partial_i u_j)$,
  and summation over repeated indices is implied.  In the classical
linear theory of elasticity, the vector $u_i$ is the displacement
field describing the physical deformation of a continuous
medium.  For usual elastic bodies, the antisymmetric combination
  $\partial_j u_i-\partial_i u_j$ corresponds to a local rotation of
the material, which is not allowed here. For granular materials, on the
  other hand, grains might locally rotate due to the presence of
  friction.  This extension which suggests a continuum description in
  terms of Cosserat elasticity was recently discussed in \cite{Gay}.
The absence of internal torques requires that the stress
tensor is also symmetric. The coefficients $\lambda_{ijkl}$ are
material constants and form the elastic modulus tensor.  The indices
$i,j,k,l$ are equal to $x,z$ where for later purposes $x$ is to be
considered as the horizontal coordinate and $z$ a vertical coordinate
pointing downward.
  
Symmetry of both the stress and the strain tensor imply a permutation
symmetry within the first and second pair of indices for $\lambda_{ijkl}$, i.e.
\beq
\label{sym}
\lambda_{ijkl}=\lambda_{jikl}=\lambda_{ijlk}=\lambda_{jilk}.  
\eeq
Materials whose behavior is modeled only in terms of equation
(\ref{s.u}) without any reference to a free energy functional are
characterized by an elastic modulus tensor that need not have any
  symmetries other than equation (\ref{sym}).  They are called
`hypoelastic' when $u_{ij}$ corresponds to a real strain tensor
\cite{truesdell}.  In hyperelastic materials, on the other hand, the
existence of quadratic free energy functional 
\beq
F=\frac{1}{2}\lambda_{ijkl}u_{ij}u_{kl} 
\eeq 
gives an additional symmetry under exchange of the first and second pair of indices, i.e.  
\beq
\lambda_{ijkl}=\lambda_{klij}.  
\eeq

In the ``pseudo-elasticity'' theory, the vector $u_i$ will be 
a novel geometric quantity (see below), and the resulting tensor $u_{ij}$
will be called a pseudo-elastic strain tensor.  This tensor is still symmetric,
as explained in section \ref{dfcn}, but the above additional symmetry is
in general {\it not} present. 

We wish to construct general solutions of the equilibrium equations
\beq
\label{equil}
\partial_i \sigma_{ij}=0.
\eeq
In order to close the problem for the stress tensor, a supplementary
condition is needed which is the condition of compatibility,
\beq
\label{compa}
\partial_z^2 u_{xx}+\partial_x^2 u_{zz}-2\partial_x\partial_z u_{xz}=0,
\eeq
resulting  simply from
the fact that the tensor $u_{ij}$ is built with the derivatives of a vector
$u_i$. This relation does not depend on a specific interpretation of the
tensor in terms of real deformations.

The entries of the stress and strain tensors can be arranged in
vector form, i.e. ${\bf \Sigma}=(\sigma_{xx},\sigma_{zz},\sigma_{xz})^T$
and ${\bf U}=(u_{xx},u_{zz},u_{xz})^T$, giving a matrix representation of
the elastic modulus tensor,
\beq
\label{stressstrain}
{\bf \Sigma}={\Lambda}\,{\bf U},
\eeq
where
\beqa
{\Lambda}=
\left(
\begin{array}{ccc}
\lambda_{xxxx} & \lambda_{xxzz} & 2\lambda_{xxxz} \\
\lambda_{zzxx} & \lambda_{zzzz} & 2\lambda_{zzxz} \\
\lambda_{xzxx} & \lambda_{xzzz} & 2\lambda_{xzxz}
\end{array}
\right).
\eeqa
The factors of $2$ are due to the symmetry under exchange of the last 2 indices of
$\lambda_{ijkl}$ and $u_{kl}$.
Now, we want to express the compatibility relation in terms of the stress
tensor, so we need to express ${\bf U}$ in terms of ${\bf \Sigma}$, i.e.
\beq
{\bf U}={\cal B}\,{\bf \Sigma},
\eeq
where ${\cal B}=(B_{ij})={\Lambda}^{-1}$. Then equation (\ref{compa}) for an
anisotropic medium is rewritten as follows
\beq
\label{compa.an}
  B_{1j}\partial_z^2\Sigma_{j} +
  B_{2j}\partial_x^2\Sigma_{j} -
2 B_{3j}\partial_x\partial_z\Sigma_{j} = 0.
\eeq
For an isotropic medium, $B_{11}=B_{22}$, $B_{21}=B_{12}$,
$B_{3i}=B_{i3}=0$, for $i=1,2$, thus the equation reduces to
$\Delta(\sigma_{xx}+\sigma_{zz})=0$. 

In the following, we will look for
solutions  of the form $\sigma_{ij}\propto e^{iqx+i\omega z}$.
In this case, equation (\ref{compa.an}), together with the conditions of
mechanical equilibrium (\ref{equil}), can be rewritten in matrix form :
\beq
{\cal A}(q,\omega){\bf \Sigma}=0.
\eeq
A non-trivial solution occurs if $\det({\cal A}(q,\omega))=0$, which leads to a certain 
{\it dispersion relation}  of the form $\omega(q)=Xq$ where $X$ obeys the 
following equation:
\beq
\label{roots}
\frac{B_{22}}{B_{11}} -
\frac{B_{23}+2B_{32}}{B_{11}} X +
\frac{2B_{33}+B_{21}+B_{12}}{B_{11}} X^2 -
\frac{B_{13}+2B_{31}}{B_{11}} X^3 + X^4 = 0.
\eeq
Depending on whether the roots $X$ are real or
complex, the response function will be qualitatively different:
\begin{itemize}

\item Complex roots, corresponding to elliptic equations for the stress, 
appear within the classical theory of anisotropic elasticity.
The fact that the roots are complex follows from the positivity of the
free energy \cite{fichera}. 

\item Purely real roots can occur in the context of directed force 
chain networks considered below. 
The existence of at least one purely real root of the dispersion relation 
classifies the problem at hand as `hyperbolic' \cite{fichera}. 

\end{itemize}

\subsection{The case of uniaxial symmetry}

Let us consider the case of uniaxial anisotropy and choose
$x$ and $z$ to be along the principal axes of anisotropy.  Then only $\lambda_{ijkl}$ with 
even numbers of equal indices is nonzero. Due to the symmetry (\ref{sym})
of $\lambda_{ijkl}$, this leaves one in general with 5 different 
constants. The matrix $\Lambda$ takes the form
\beqa
\label{pseudo.ela}
\Lambda_\dagger=
\left ( \bay{ccc} a & c & 0 \\
                  c' & b & 0 \\
                  0 & 0 & d          \eay \right ).
\eeqa
We denote
it with a dagger to indicate that it corresponds to a material with a vertical uniaxial
symmetry.  An alternative parametrization of $\Lambda_\dagger$, standard in elasticity theory, 
is
\be
\Lambda_\dagger = \frac{1}{1-\nu_x\nu_z}\left (
\bay{ccc}
E_x       & \nu_z E_x & 0 \\
\nu_x E_z & E_z       & 0 \\
0         & 0         & (1-\nu_x\nu_z)G
\eay
\right ),
\ee
where $E_{x,z}$ and $G$ are the Young and shear moduli respectively, 
and $\nu_{x,z}$ the Poisson ratios.
Note that the present form includes a linear elasticity theory without a
free energy functional. The classical theory is recovered with the extra symmetry $c'=c$.
In this case, $E_x$, $E_z$, $\nu_x$ and
$\nu_z$ are not independent, satisfying the relation
$\frac{E_z}{E_x} = \frac{\nu_z}{\nu_x}$. Together with $G$, we are
thus left with four independent constants.

In classical elasticity theory for a uniaxial system, the stress-strain relation is derivable from 
an energy density of the form
\be
F = \frac{1}{2}\left[ a\uxx^2 + b\uzz^2 + 2c\uxx\uzz + 2d\uxz^2\right],
\ee
The material described is stable under deformations if and only if
$F$ is positive definite for any strain, which requires
\be
a>0; \quad b>0; \quad d>0; \quad \mbox{and} \quad ab-c^2>0.
\ee 
Or, equivalently,
\be \label{eq:stability}
\nu_x \nu_z < 1; \quad E_x>0; \quad E_z>0; \quad \mbox{and} \quad G>0.
\ee 
An elastic material that is permitted to reversibly deform must obey these
constraints, but they do not apply to materials for which there is no
well-defined free energy quadratic in the strains.  We speak of such
materials as being described by coefficients that lie outside the 
``classical stability'' range.

The compatibility condition (\ref{compa}) expressed in terms of the stress tensor reads:
\beq
b\dr^2_z\sxx - c \dr^2_z\szz - c'\dr^2_x\sxx + a\dr^2_x\szz
-2 \frac{\det \Lambda}{d^2} \dr_x\dr_z\sxz = 0
\label{compat2}
\eeq
Combining this relation with the two equilibrium conditions of equation~(\ref{equil}),
\bea
\dr_z\szz + \dr_x\sxz & = & 0, \label{equiz} \\
\dr_z\sxz + \dr_x\sxx & = & 0, \label{equix}
\eea
we obtain, for any one of the components of the stress tensor:
\be
\label{sijequation}
\left ( \dr^4_z + t\dr^4_x + 2 r \dr^2_x\dr^2_z \right ) \sij =  0,
\ee
where the coefficients $t$ and $r$ are given by
\bea
\label{r.t}
t & = & \frac{a}{b} = \frac{E_x}{E_z}, \nonumber\\
r & = & \frac{ab-cc'-\frac{1}{2}d(c+c')}{bd} = 
\frac{1}{2}E_x \left( \frac{2}{G} - \frac{\nu_z}{E_z}-\frac{\nu_x}{E_x} \right ).
\eea

Expanding the stresses in Fourier modes, it is easy to see that the
solutions of the equations (\ref{equiz}-\ref{sijequation}) are of the
form
\bea
\szz & = &          \iqt \sum_{k} a_k(q) \, e^{iqx + iX_kqz},
\label{szzFourier} \\
\sxz & = & C_{xz} - \iqt \sum_{k} a_k(q) \, X_k   \, e^{iqx + iX_kqz},
\label{sxzFourier} \\
\sxx & = & C_{xx} + \iqt \sum_{k} a_k(q) \, X_k^2 \, e^{iqx + iX_kqz},
\label{sxxFourier}
\eea
where $C_{xx}$ and $C_{xz}$ are constants. From equation (\ref{sijequation})
we see that the $X_k$ are the roots of the following quartic equation
\be
X^4 + 2 r X^2 + t = 0,
\label{equaX}
\ee
a special case of equation~(\ref{roots}).  There are four solutions:
\be \label{rootX}
X = \pm \sqrt{- r \pm \sqrt{r^2 - t}}.
\ee
Hence the index $k$ runs from $1$ to $4$.  The four functions $a_k(q)$
and the constants $C_{xx}$ and $C_{xz}$ must be determined by the boundary
conditions, as shown in section \ref{response} and Appendix B.

We see that only two combinations, $r$ and $t$, of the 5
elastic constants will determine the structure of the response
function in anisotropic materials.  

\subsection{Main results of this paper}

\begin{figure}[tpb]
\begin{center} \epsfxsize=10cm
\epsfig{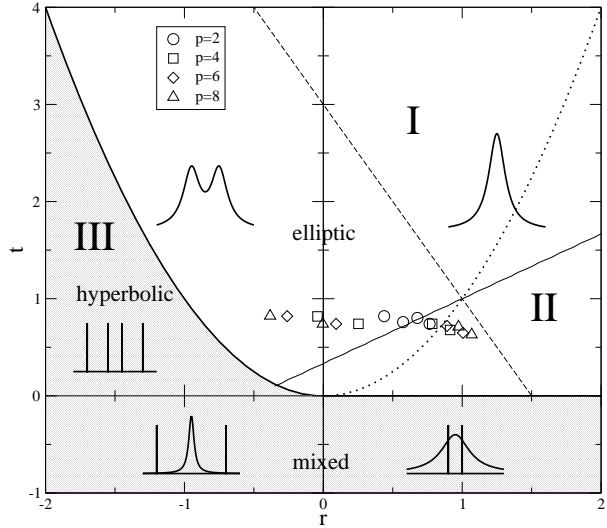}
\end{center}
\caption{$(r,t)$ phase diagram characterizing the qualitative nature of the
stress profiles. The shaded region corresponds to hyperbolic and ``mixed'' equations for
stresses whereas the unshaded region allows for elliptic equations. The hyperbolic region is bounded above by the line $t=r^2$,
separating it from the elliptic region.
 In the elliptic
region, a double peak stress profile
is found in the whole region $r<0$. The solid and dashed
straight lines are the trajectories for the triangular spring network 
studied in section \ref{triang}, for horizontal and vertical orientation of one of the springs,
respectively. 
The symbols correspond to the solutions of the anisotropic linear {\sc dfcn} model for various values of 
the anisotropic scattering parameter $p$, see section \ref{dfcn}.
}
\label{fig0}
\end{figure}

We show in figure \ref{fig0} the various `phases' in the $r$-$t$ plane corresponding to different shapes
of the response function, as obtained from the calculation 
presented in section \ref{response} below. 

The line $t=r^2$, for $r<0$,
separates the hyperbolic and the elliptic regions. For $t > r^2$ (region I), 
the above roots $X_k$ are complex and
we write:
\bea
X_1 = - X_4 = \beta - i \alpha,  \label{X1and4caseI} \\
X_2 = - X_3 = -\beta - i \alpha, \label{X2and3caseI}
\eea
where $\alpha$ and $\beta$ are positive real numbers. When $t < r^2$, 
$r >0$ (region II), one the other hand,
the roots $X_k$ are purely imaginary and one has:
\bea
X_1 = - X_4 = - i \alpha_1, \label{X1and4caseII} \\
X_2 = - X_3 = - i \alpha_2, \label{X2and3caseII}
\eea
where $\alpha_1$ and $\alpha_2$ are positive real numbers. 

Note that the isotropic limit corresponds to the point $r=1,\, t=1$. As
we show in detail in section \ref{response}, the elliptic region
contains a subregion $r<0$, $t > r^2$, where the response function has
a two peak structure with peak widths growing linearly with depth. As one
approaches the line $t=r^2$, the two peaks become narrower and
narrower, finally becoming two delta-function peaks exactly on the
transition line.  Below the transition, there is a hyperbolic regime
(region III in figure \ref{fig0}) where the response consists of four
delta-function peaks.

The parameter range $t<0$, labeled ``mixed'' in figure \ref{fig0}, gives rise
to a third type of behavior of the response function due to the fact
that there are two real roots and two imaginary roots. It may only
appear in the non-stable pseudo-elastic case, and gives superposition
of a hyperbolic two delta peak response function and a single-peka classically elastic response function. For the particular model
for the {\sc DFCN} discussed below, the range $t<0$ does not
occur. Hence this case is not pursued any further here.

We discuss below some particular trajectories in the $r$-$t$ plane
(see sections \ref{triang} and \ref{dfcn}).
One corresponds to simple, anisotropic, triangular networks of springs,
that lead on large scales to classical anisotropic elasticity with parameters 
on the plain and dotted straight lines, corresponding to two orientations
of the lattice (see figure \ref{figsprings}).  Both trajectories meet
at the point $(1,1)$ corresponding to an isotropic medium where all
springs have the same stiffness. Moreover, both trajectories cross the
region $r<0$ and thus allow for two peak response functions.
Inclusion of three-body forces permits spring networks with $(r,t)$
anywhere in region I or II (see section \ref{triang}).

We have also computed $r$ and $t$ for the linear {\sc dfcn} model, for
a particular model for scattering where the degree of anisotropy is
tuned in terms of a parameter $p$ (see section \ref{dfcn}). The results
are shown as symbols, and appear to always lie in the elliptic region.
As in the spring networks, for sufficiently anisotropic scattering, one enters the
region $r < 0$ where the response function has two peaks.

In two classical papers \cite{green}, Green et al. have treated the stress distribution inside
plates with two directions of symmetry with right angles to each other
. The solutions are parametrized, apart from boundary
conditions, by $\alpha_1, \alpha_2$ (not to be confused with
$\alpha_i$ introduced above) which are related to the set $r,t$ by
$(r+\sqrt{r^2-t}/t),(r-\sqrt{r^2-t}/t) $. The authors assume their parameters
$\alpha_1, \alpha_2$ to be always real and positive, based on
empirical fits of elastic constants for timbers
such as oak and spruce. This choice corresponds to region II in figure
\ref{fig0}. Consequently, the possibility of region I and III behavior, and
particularly the appearance of a double peak response for a
classically elastic material, is not discussed in
\cite{green}. Moreover, their analysis considers the response
in the case where the boundaries and the directions of symmetry are
either parallel or perpendicular to each other, whereas the present
discussion - see in particular section III.B - treats a more general case. The
response functions for region II, as computed in the present work,
could in principle be reconstructed from the results of \cite{green}.

\section{Shape of the response function}
\label{response}

After having discussed the general framework of anisotropic elasticity
and the particular example of two-dimensional systems with uniaxial symmetry, we now turn
to the actual shape of the response function in such materials.
We will calculate the response of an elastic or pseudo-elastic slab of infinite horizontal
extent to a localized force applied at the top surface. We shall consider
the case of a semi-infinite system with a force applied at a single point on
its surface, for which complete analytical solutions can be obtained. More
general situations (finite spatial extension of the overload, finite thickness
of the slab with a rough or smooth bottom, ...) should be considered to obtain 
quantitative fits of experimental \cite{Reydellet,Reydelletbis} and numerical
data. Still, two angles are left free: the angle
$\theta_0$ that the applied force makes with the vertical, and the
orientation angle $\tau$ of the anisotropy with the vertical.

\subsection{Vertical anisotropy}

We are interested in the response of a semi-infinite system to a localized
force at its top surface $z=0$. We suppose that this force is of amplitude
$F_0$ and makes an angle $\theta_0$ with the vertical direction, as shown in
figure \ref{figF0}.

\bfig[b!]
\bc
\epsfxsize=4cm
\epsfbox{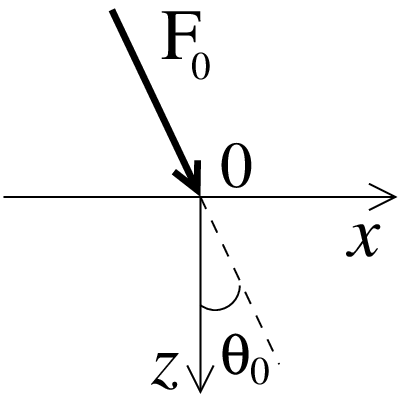}
\caption{Force at the top surface.
\label{figF0}}
\ec
\efig

The corresponding stresses at $z=0$ are then
\bea
\szz & = & F_0 \cz \, \delta(x), \label{zztop} \\
\sxz & = & F_0 \sz \, \delta(x). \label{xztop}
\eea
To obtain the results described below, we make use of the identity
\be
\delta(x) = \frac{1}{2\pi} \iqz (e^{iqx} + e^{-iqx}),
\ee
and impose the boundary condition by identifying the coefficients 
of $e^{\pm iqx}$ in the equations (\ref{zztop}-\ref{xztop}) 
and (\ref{szzFourier})-(\ref{sxxFourier}) at $z=0$.
Note that $\sxx(z=0)$ is not determined by the boundary conditions.

When $z \to +\infty$, we expect all stresses to decay to zero.
It turns out that this is a self-consistent condition as long
as the system is energetically stable, but cannot be imposed
in the unstable regime. The
reader interested  in a more detailed derivation of the following results
can consult appendix B.

\bfig[t!]
\bc
\epsfxsize=7cm
\epsfbox{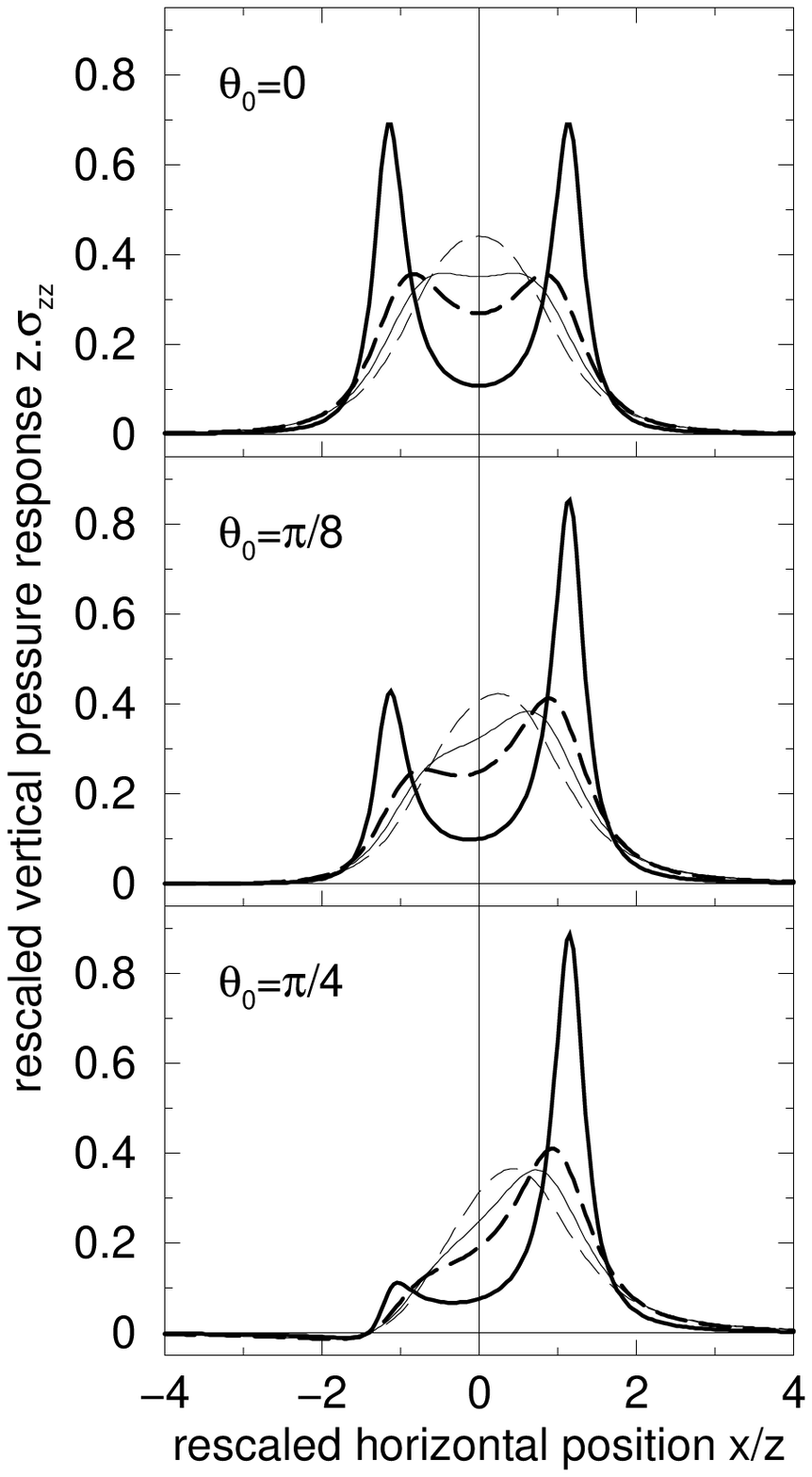}
\caption{{\it Region I}: Rescaled stress profiles for several directions $\theta_0$ of
the applied force and several values of $r$, with $t=2$.
In each panel, the thick solid line is for $r=-1.3$, the thick
dashed line is for $r=-0.7$, and the thin solid line is
for $r=-0.2$, and the thin dashed line is for $r=0.5$. $r>0$ is the condition to have a single
peaked profile for $\theta_0=0$.
\label{figcaseI}}
\ec
\efig

\subsubsection*{Region I (elliptic): $t > r^2$}

Since we want all the stresses to vanish at large depth, the functions $a_1$ and $a_2$ in 
(\ref{szzFourier})-(\ref{sxxFourier}) must be
zero for $q>0$, and $a_3$ and $a_4$ must vanish for $q<0$. In addition,
$C_{xx}$, and $C_{xz}$ must all vanish. Furthermore, because the stresses
are real quantities, $a_1(-q)=a^*_3(q)$ and $a_2(-q)=a^*_4(q)$. 

The boundary conditions at $z=0$ then imply
\bea
a_3 & = & \frac{F_0}{4\pi\beta}\left[(\beta-i\alpha)\cz - \sz\right], \\
a_4 & = & \frac{F_0}{4\pi\beta}\left[(\beta+i\alpha)\cz + \sz\right].
\eea
Since the coefficients $a_3$ and $a_4$ are independent of $q$, the integrals
in equations (\ref{szzFourier})-(\ref{sxxFourier}) are straightforwardly
carried out, yielding
\bea
\szz & = & \frac{F_0}{2\pi} \,
           \frac{4\alpha z^2 [ z\cz (\alpha^2+\beta^2) + x\sz]}
                {[(\alpha^2-\beta^2)z^2 + x^2]^2 + [2\alpha\beta z^2]^2}, \\
\sxz & = & \frac{x}{z}\,\szz, \\
\sxx & = & \left(\frac{x}{z}\right)^2 \szz.
\eea
The latter two results follow directly from the observation that the
integrals in equations (\ref{sxzFourier}) and (\ref{sxxFourier}) can
be expressed simply as convolutions of $\szz(q)$ with the Fourier
transforms of $x/z$ and $x^2/z^2$, respectively. In the limit $\beta\to 0$ 
(which corresponds to $r^2-t\to 0$) and $\alpha \to 1$, we recover the
familiar isotropic formulas \cite{LL}.

Figure \ref{figcaseI} shows the response for four different choices
of the parameter $r$ and a fixed $t$, each being shown for three choices of
$\theta_0$. Note that $\szz$ has a more pronounced double-peak
structure for increasingly negative $r$. For $\theta_0=0$, the
condition for having a double peak is $\dr_x^2\szz(x=0)>0$, which
occurs when $\alpha^2<\beta^2$, or equivalently $r<0$. In
terms of the Young and shear moduli and the Poisson ratios, this condition
can be expressed as $G > E_x/\nu_x = E_z/\nu_z$. The
positions of the peaks are then given by $x=\pm z
\sqrt{\beta^2-\alpha^2}= \pm z \sqrt{|r|}$. From the curvature at the maximum, one can 
define a width $w$ of these peaks which reads: 
\be
w = \frac{\alpha \beta}{\sqrt{2}}\frac{1}{\sqrt{\beta^2-\alpha^2}} z
  = \frac{\sqrt{t-r^2}}{2 \sqrt{2|r|}} z.
\ee
Thus the peaks become sharper and sharper as one approaches the hyperbolic limit
$t=r^2$. 

A very important point is that the response profiles scale with the reduced
variable $x/z$ when multiplied by the height $z$. This means that,
when the profile is double peaked, these two peaks get larger in the same
way that they get away from each other. Such a response cannot therefore
be seen as an `hyperbolic-like' signature, for which the peak width compared
to the distance between the peaks goes to zero at large depth.
However, in the limit where $t \to r^2$, the width of the peak vanishes a 
the response becomes truly hyperbolic.

\bfig[t!]
\bc
\epsfxsize=7cm
\epsfbox{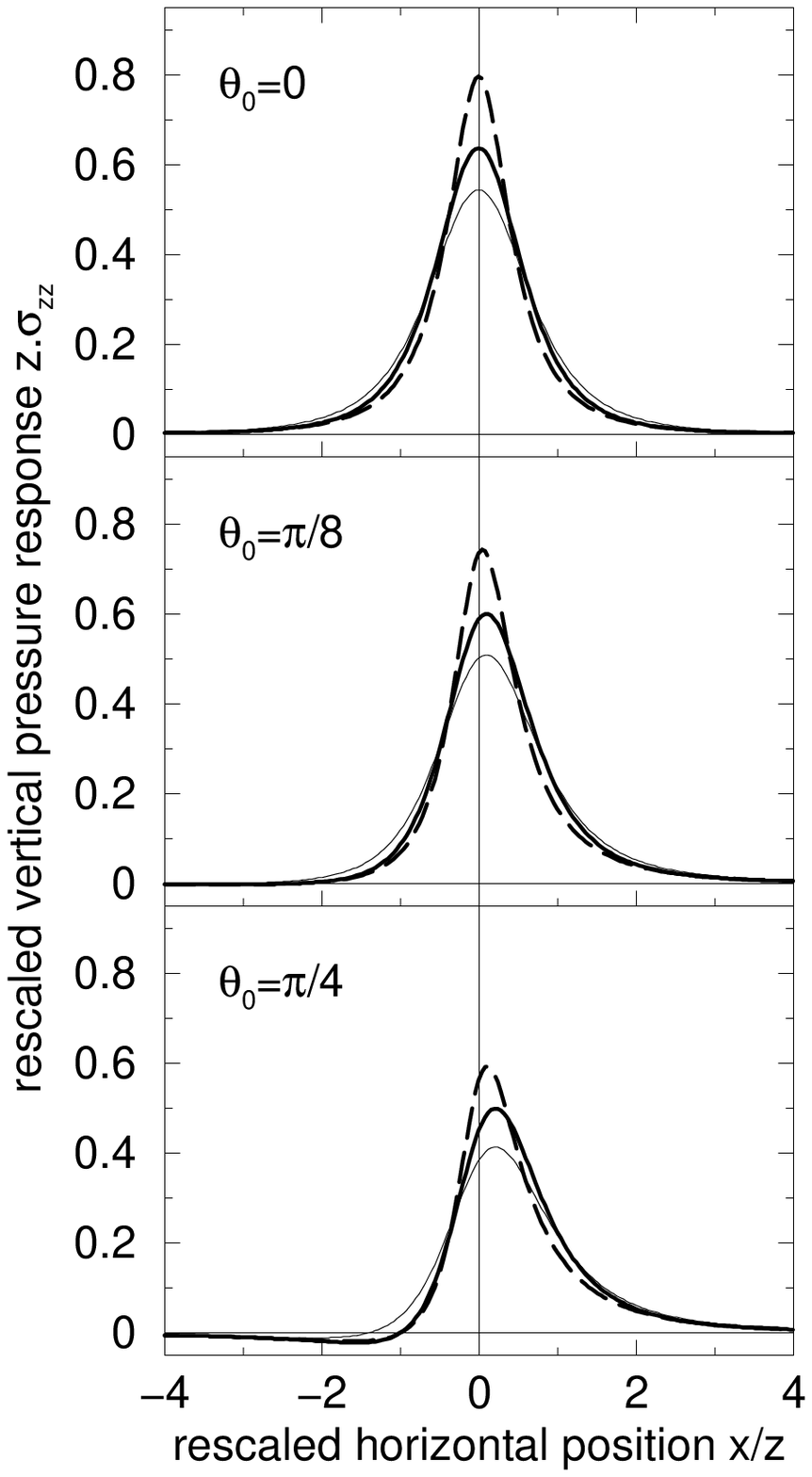}
\caption{{\it Region II}. Stress profile for different cases.
The solid thick line is for $t=1$ and $r=1$ (isotropic case),
the thick dashed line is for $t=1$ and $r=2.125$, and the solid thin line is
for $t=2$ and $r=1.5$.
\label{figcaseII}}
\ec
\efig

\subsubsection*{Region II (elliptic): $t < r^2$, $r >0$}
Again, we 
only keep the functions $a_1$ and $a_2$ for $q<0$, and $a_3$ and $a_4$ for
$q>0$. This time, the fact that stresses are real quantities requires 
$a_1^*(-q)=a_4(q)$ and $a_2^*(-q)=a_3(q)$.
A similar analysis to the above yields
\bea
\szz & = & \frac{F_0}{2\pi} \, 
           \frac{2(\alpha_1+\alpha_2) z^2 [\alpha_1\alpha_2 z \cz + x \sz]}
           {[(\alpha_1 z)^2 + x^2][(\alpha_2 z)^2 + x^2]}, \\
\sxz & = & \frac{x}{z} \, \szz, \\
\sxx & = & \left(\frac{x}{z}\right)^2 \szz.
\eea
For $\alpha_1=\alpha_2=1$ (again $r^2-t = 0$), we recover the isotropic
formula. In this case, however, when $\theta_0=0$, $\szz$ always presents a 
single peak, see figure \ref{figcaseII}. Depending on the values of
$\alpha_1$ and $\alpha_2$, the profiles can be broader or narrower than the
isotropic response, as has been observed experimentally on, respectively, 
dense and loose packings \cite{Reydellet}.

\subsubsection{Region III (hyperbolic): $t < r^2$, $r  < 0$}

In this case, all the roots $X_k$ are real, and the response function is 
the sum of four $\delta$ peaks, at positions $x = X_k z$. 
The appearance of four peaks is different from previous hyperbolic
models \cite{us,Proc} giving two peaks in which case the closure relation for the
stresses is linear, whereas here the closure is achieved by a 4th
order partial differential equation, Eq.(\ref{sijequation}).  
The four 
peaks merge into two peaks exactly on the hyperbolic-elliptic boundary
$t=r^2$. 
The reason why previous hyperbolic models \cite{us,Proc} work so well
could be that granular system such as sandpiles are close to the
hyperbolic-elliptic boundary (see also section IV.B for further remarks).
Inside region III, the fact that all roots are real excludes
the possibility to require stresses to vanish for large
$z$. This leads to a situation where there are more constants of
integration than boundary conditions.

One may advance on the analytical form of response functions
using physical arguments as follows.
Let us first rewrite the equation for stresses (\ref{sijequation}), as follows
\be
\label{sijequation.III}
(\partial_z^2-c_+^2\partial_x^2)(\partial_z^2-c_-^2\partial_x^2)\sigma_{ij}=0
\ee
where 
\be
c_\pm^2=-r\pm\sqrt{r^2-t}
\ee
leading to $c_\pm\geq 0$. The constants $\pm c_\pm$ are just the four
real roots $X_k$ mentioned above. Instead of solving the equation above, we
consider special solutions $\sigma_{ij}^+$, $\sigma_{ij}^-$ of the
following PDE,
\be
(\partial_z^2-c_\pm^2\partial_x^2)\sigma_{ij}^\pm=0,
\ee
which automatically satisfy equation (\ref{sijequation.III}).
Both equations can be solved for the boundary conditions
(\ref{zztop}) and (\ref{xztop}), giving the solutions
\bea
\label{sol.spec.III}
\szz^\pm & = & \frac{F_0}{2} \, 
           \left(
\left[\cz -\frac{\sz}{c_\pm}\right]\delta(x+c_\pm z)
+\left[\cz +\frac{\sz}{c_\pm}\right]\delta(x-c_\pm z)
\right), \\
\sxz^\pm & = & \frac{F_0}{2} \, 
           \left(
-\left[c_\pm\cz -\sz\right]\delta(x+c_\pm z)
+\left[c_\pm\cz +\sz\right]\delta(x-c_\pm z)
\right), \\
\sxx^\pm & = & 
\frac{F_0}{2} \, 
           \left(c_\pm
\left[c_\pm\cz -\sz\right]\delta(x+c_\pm z)
+c_\pm\left[c_\pm\cz +\sz\right]\delta(x-c_\pm z)
\right)
.
\eea
Before constructing a general solution from $\sij^\pm$, let us remark
that there are in principle additional solutions $\tilde{\sigma}_{ij}$
satisfying 
\be
(\partial_z^2-c_\pm^2\partial_x^2)\tilde{\sigma}_{ij}=\sij^\mp
\ee
However, these solutions are not finite as they involve divergences
arising from integrals such as $\int_{-\infty}^\infty dq \cos (qu)/q^2$ . Therefore, we conclude that
a general solution of equation (\ref{sijequation.III}) may be 
constructed as
\be
\label{sol.III}
\sij=a_+\sij^+ +a_-\sij^-.
\ee
It should satisfy the boundary conditions
(\ref{zztop}) and (\ref{xztop}) which yield a relation
\be
a_+ +a_-=1.
\ee
The coefficients $a_+$ and $a_-=1-a_+$ are relative weights which
indicate how the applied load is shared between
the two sets of force chains characterized by $c_\pm$. As there is no
physical mechanism introduced a priori which
prefers one set of force chains to the other, we are left with one
free parameter, say $a_+$, for the response function $\sij$.  
The ambiguity on the value of $a_+$ could be resolved by considering
e.g. a 
microscopic model that leads to equation
(\ref{sijequation.III}). 
\bfig[t!]
\bc
\epsfxsize=8cm
\epsfbox{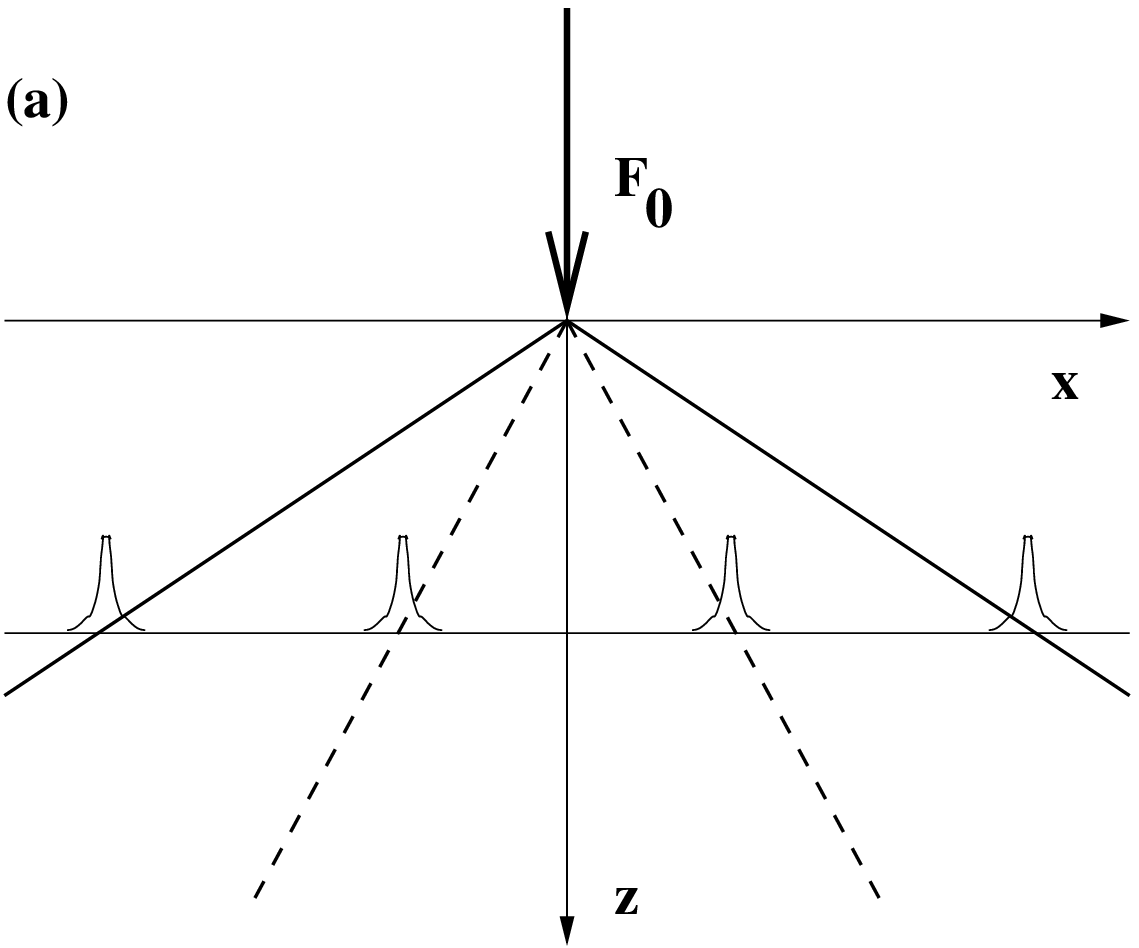}
\epsfxsize=8cm
\epsfbox{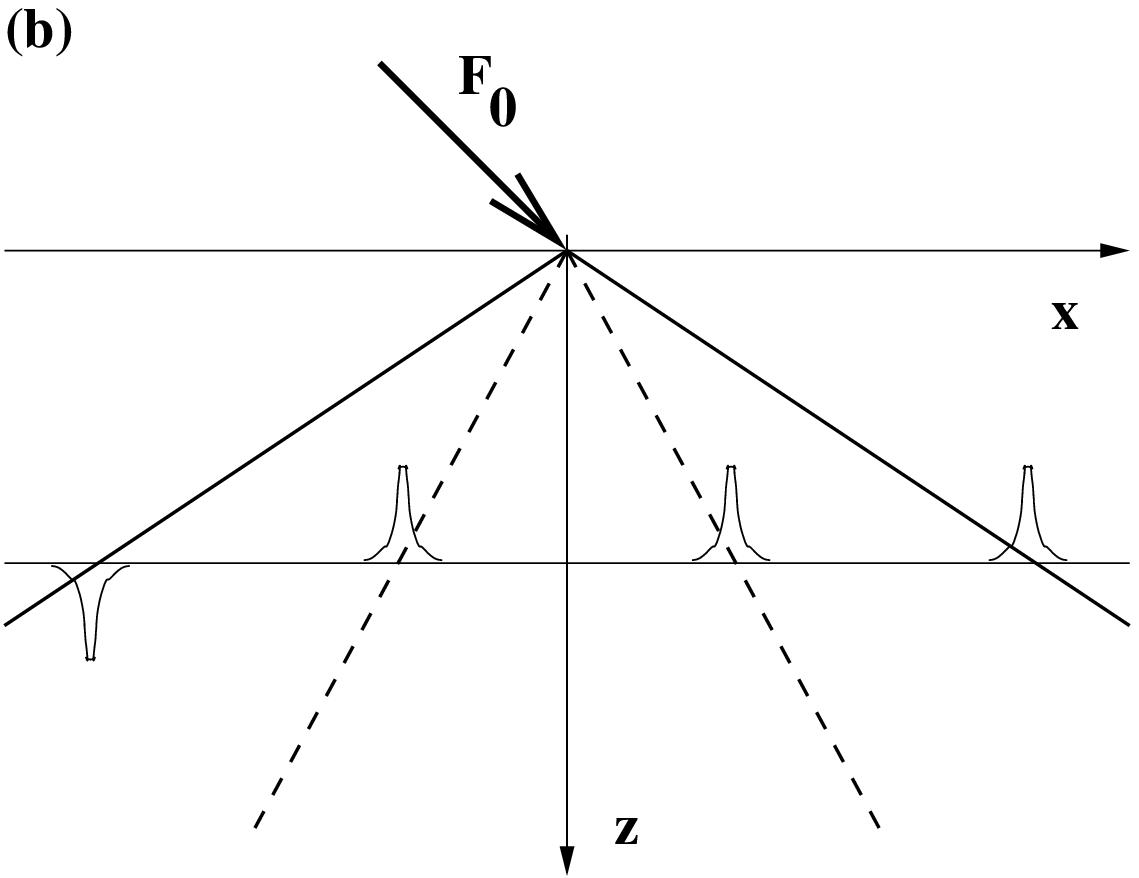}
\caption{{\it Region III}. The stress profile is a sum of four delta
  functions. The characteristics $x=\pm c_\pm z$ along which 
the applied load is propagated are shown. Parameters are $r=-1.0$, $t=0.75$
giving $c_+=1.5$ (solid lines) and  $c_-=0.5$ (dashed lines). The
delta functions are indicated by cartoons. (a)
$\theta_0=0$, (b) $\theta_0=\pi/4$.
\label{figcaseIII}}
\ec
\efig

In figure \ref{figcaseIII}, the propagation of the applied force along
the characteristics is shown. Note that the sign of $\szz$ may change
along a certain characteristic
if $\cz -\frac{\sz}{c\pm}<0$ (see figure \ref{figcaseIII}-(b)). 

\subsection{Anisotropy at an angle}
\label{angle}
We now, for completeness, generalize the results of the previous subsections to
the case where the direction of the anisotropy makes an arbitrary angle $\tau$ with
the vertical. (The previous section corresponds to $\tau=0$). 
This situation may be relevant for systems that are initially
sheared as in the experiments of Geng et al. \cite{geng}, or prepared in a way which breaks the symmetry $x\leftrightarrow -x$.
We restrict the discussion to regions I and II (the computation for
region III can be carried out in a similar fashion).

The equivalent of the relation (\ref{stressstrain}) involves now
a matrix $\Lambda_\tau$ which is related to $\Lambda_\dagger$ of equation (\ref{pseudo.ela}) by
\be\label{rotation}
\Lambda_\tau = {\cal Q}^{-1}  \Lambda_\dagger  {\cal Q},
\ee
where ${\cal Q}$ is the rotation matrix
\be
{\cal Q} = \left ( \bay{ccc}
\cos^2\tau       & \sin^2\tau        & -2\sin\tau\cos\tau \\
\sin^2\tau       & \cos^2\tau        & +2\sin\tau\cos\tau \\
\sin\tau\cos\tau & -\sin\tau\cos\tau & \cos^2\tau-\sin^2\tau
\eay \right ).
\ee

The differential equation on the stress components that is deduced
from the compatibility condition and stress-strain relations is now
much more complicated, but the corresponding roots of the fourth order
polynomial that appear when looking at Fourier modes can still be
calculated from the $X_k$ solutions of (\ref{equaX}). They read 
\be
Y_k = \frac{X_k - \tan\tau}{1 + X_k\tan\tau}, \quad k=1,\dots,4.  
\ee 
The same method as above -- see also Appendix B -- can then
be applied to find the stress response functions for a localized
overload at the top surface of the material. 
Note that the material
properties are still determined by the $X_k$ associated with $\Lambda_\dagger$.
In particular, whether the response is elliptic or hyperbolic cannot
depend on $\tau$.  In the following, regions I and II are defined with
respect to $X_k$ as above.

\subsubsection*{Region I}

The $X_k$ are of the form $\pm \beta \pm i \alpha$, see
(\ref{X1and4caseI}-\ref{X2and3caseI}). The corresponding $Y_k$ can be
constructed with the following quantities
\bea
A  & = & \frac{\alpha (1 + \tan^2\tau)}
              {(1+\beta\tan\tau)^2 + (\alpha\tan\tau)^2}, \\
B  & = & \frac{\beta (1 - \tan^2\tau) + \tan\tau (\alpha^2+\beta^2-1)}
              {(1+\beta\tan\tau)^2 + (\alpha\tan\tau)^2}, \\
A' & = & \frac{\alpha (1 + \tan^2\tau)}
              {(1-\beta\tan\tau)^2 + (\alpha\tan\tau)^2}, \\
B' & = & \frac{\beta (1 - \tan^2\tau) - \tan\tau (\alpha^2+\beta^2-1)}
              {(1-\beta\tan\tau)^2 + (\alpha\tan\tau)^2}.
\eea
The same boundary conditions -- see figure \ref{figF0} --  lead to
\bea
\szz & = & \frac{F_0}{2\pi} \,
           \frac{2 z^2}{[(x+Bz)^2 + (Az)^2][(x-B'z)^2 + (A'z)^2]} \,
           \left \{  x \sz (A+A') \right .  \\ \nonumber
     &   & \left . + z \cz [AA'(A+A')+AB'^2+A'B^2]
                   + [x \cz + z \sz] (A'B-AB') \right \}.
\eea
$\sxz$ and $\sxx$ are related to $\szz$ by the usual factors of $x/z$ and
$(x/z)^2$ respectively.

Figures \ref{t0=0caseI} and \ref{t0=taucaseI} show the pressure
response profile as different parameters are varied. In
figure \ref{t0=0caseI} the applied force is kept vertical ($\theta_0=0$),
and $\tau$ is varied from $0$ to $\pi/4$. 
Interestingly, the initially
double peaked profile (figure \ref{t0=0caseI}-(a)) is progressively
deformed in such a way that the left peak gets more pronounced,
until the remaining single peak moves to the right for
$\tau=\pi/4$. This behavior might be counter-intuitive for smaller $\tau$, because
a positive value of $\tau$ means that the main direction of
the anisotropy is oriented to the right. 
However, it
can be understood within
the ball-and-spring model of section~\ref{triang}, where the $k_1$
springs are horizontal.  Rotating to the right the two stiff
directions $k_2$ emerging from a ball downwards brings the left one closer to the vertical direction,
which therefore gets a larger fraction of the overload. 
Continuing past
 $\tau=\pi/6$, however, 
the stiffer springs form lines that slope downward to the right.
Since they continue to support most of the
load, the single peak is shifted to the right. This behavior holds
also for the single peaked profiles of figure \ref{t0=0caseI}-(b). 

The second series of plots -- figure \ref{t0=taucaseI} -- is for
the case where the applied force is exactly in the direction of the
anisotropy ($\theta_0 = \tau$). The corresponding curves are
qualitatively similar to those of Figure \ref{t0=0caseI}. 
The direction of the force imposed at the top does
not change the general shape (anisotropic double or single peak)
except for the fact that a negative pressure zone evolves for large
negative x. 

The value of $0.6$ for $t$ used in the figures
\ref{t0=0caseI} and \ref{t0=taucaseI} is motivated by experimental
findings \cite{eric}. The response function shown in figure
\ref{t0=0caseI}-(b) for $\tau=\pi/4$ is at least qualitatively
consistent with the response functions measured in \cite{geng}.
\bfig[t!] 
\bc 
\epsfxsize=0.49\linewidth 
\epsfbox{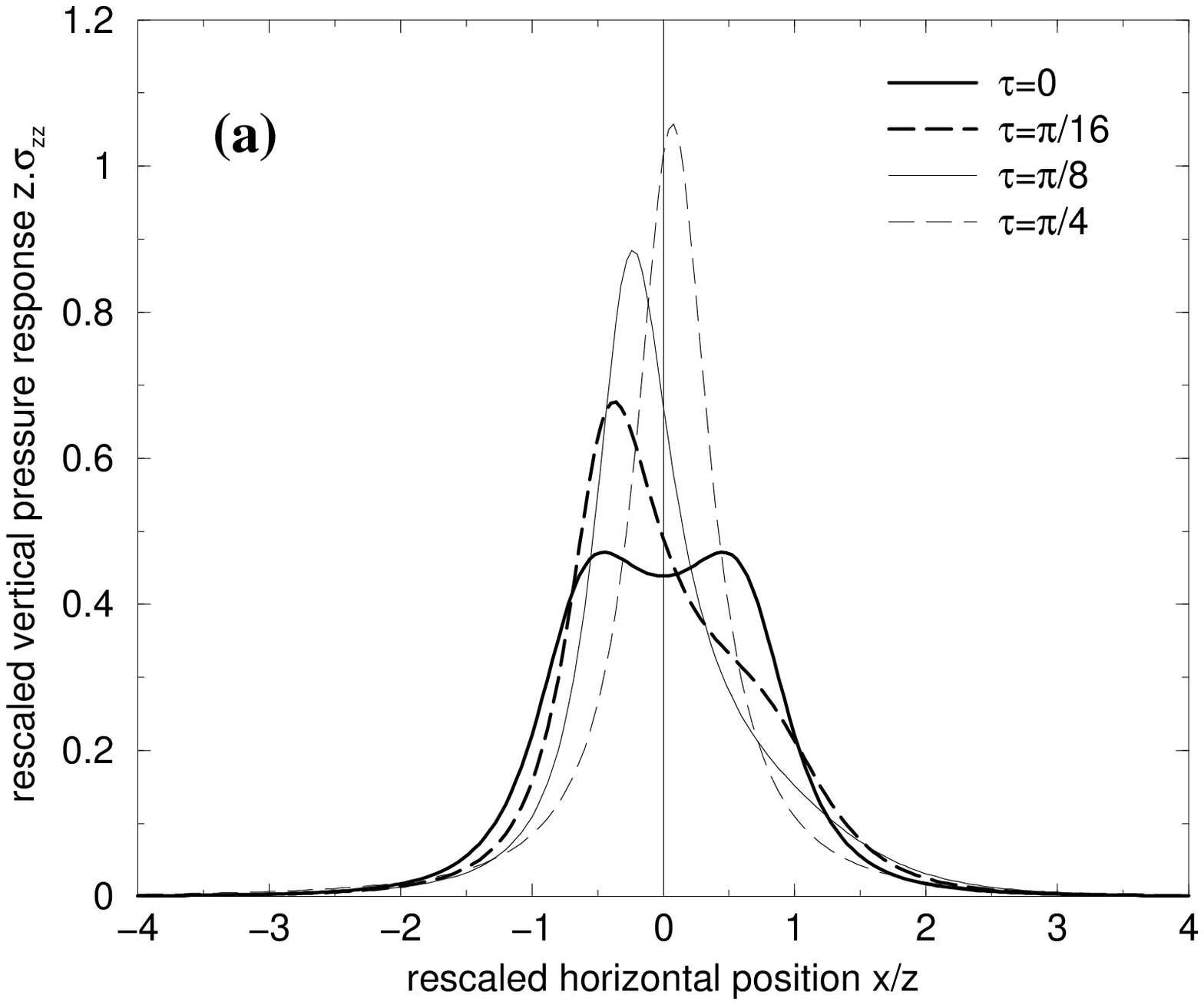} 
\hfill 
\epsfxsize=0.49\linewidth 
\epsfbox{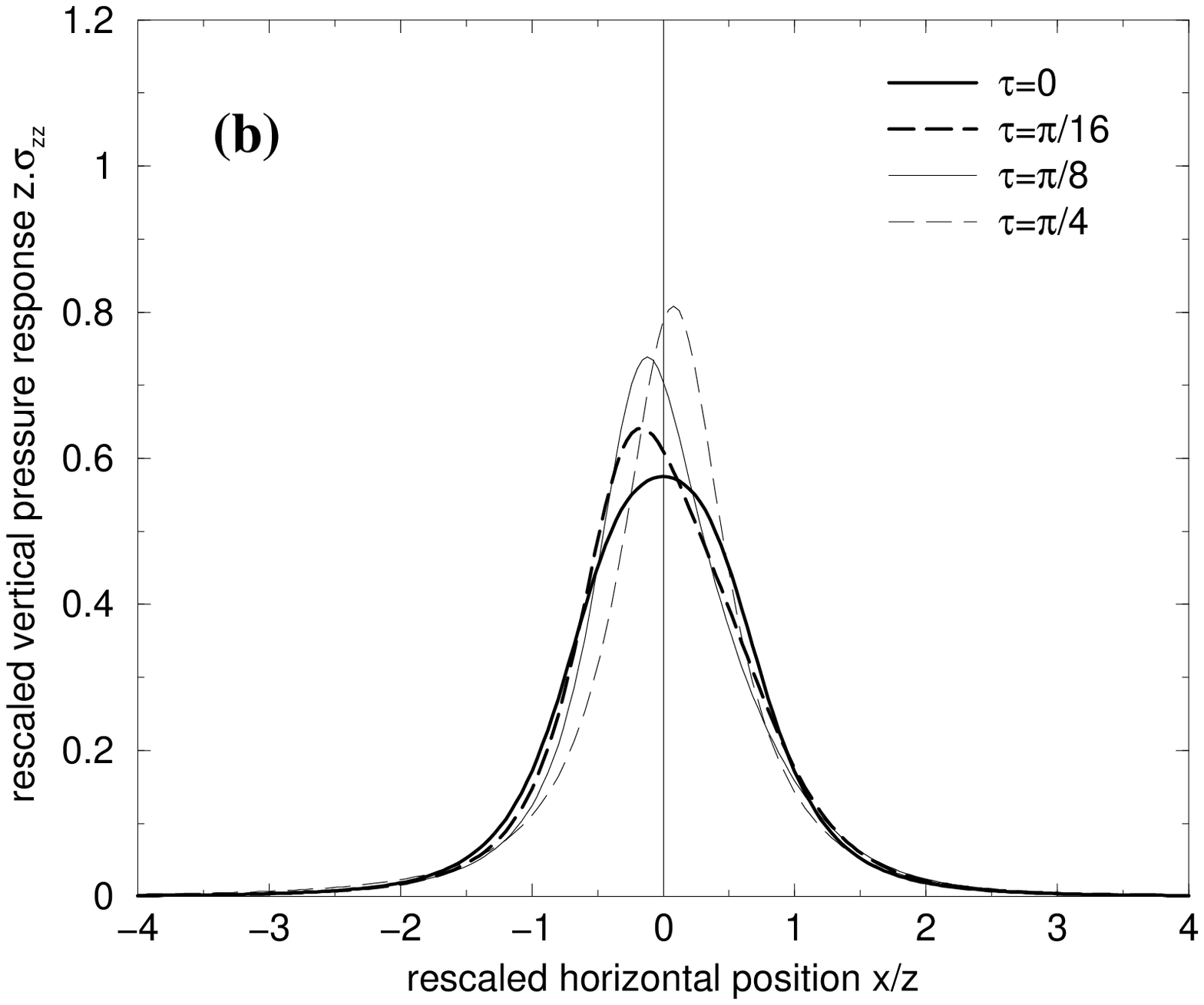} 
\caption{{\it Region I}. Response profiles for different values of the anisotropy angle 
$\tau$, but with a fixed value for the orientation of the applied force: 
$\theta_0=0$. The graph (a) is for $t=0.6$ and $r=-0.2$, while (b) has been 
obtained for $t=0.6$ and $r=0.2$. Note that for the three smallest $\tau>0$ 
the response is stronger in the negative $x$ region. 
\label{t0=0caseI}} 
\ec 
\efig 
\bfig[t!] 
\bc 
\epsfxsize=0.49\linewidth 
\epsfbox{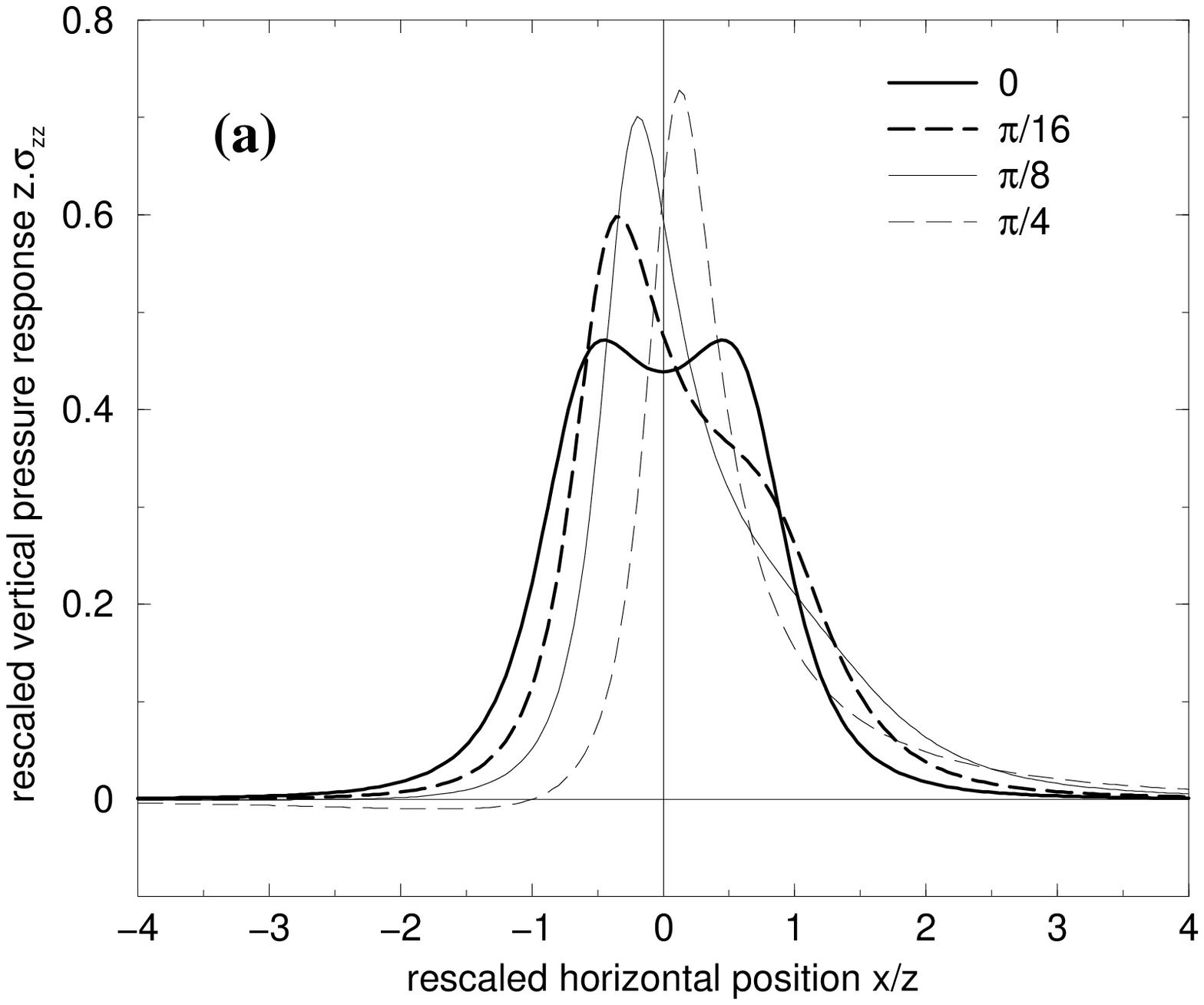} 
\hfill 
\epsfxsize=0.49\linewidth 
\epsfbox{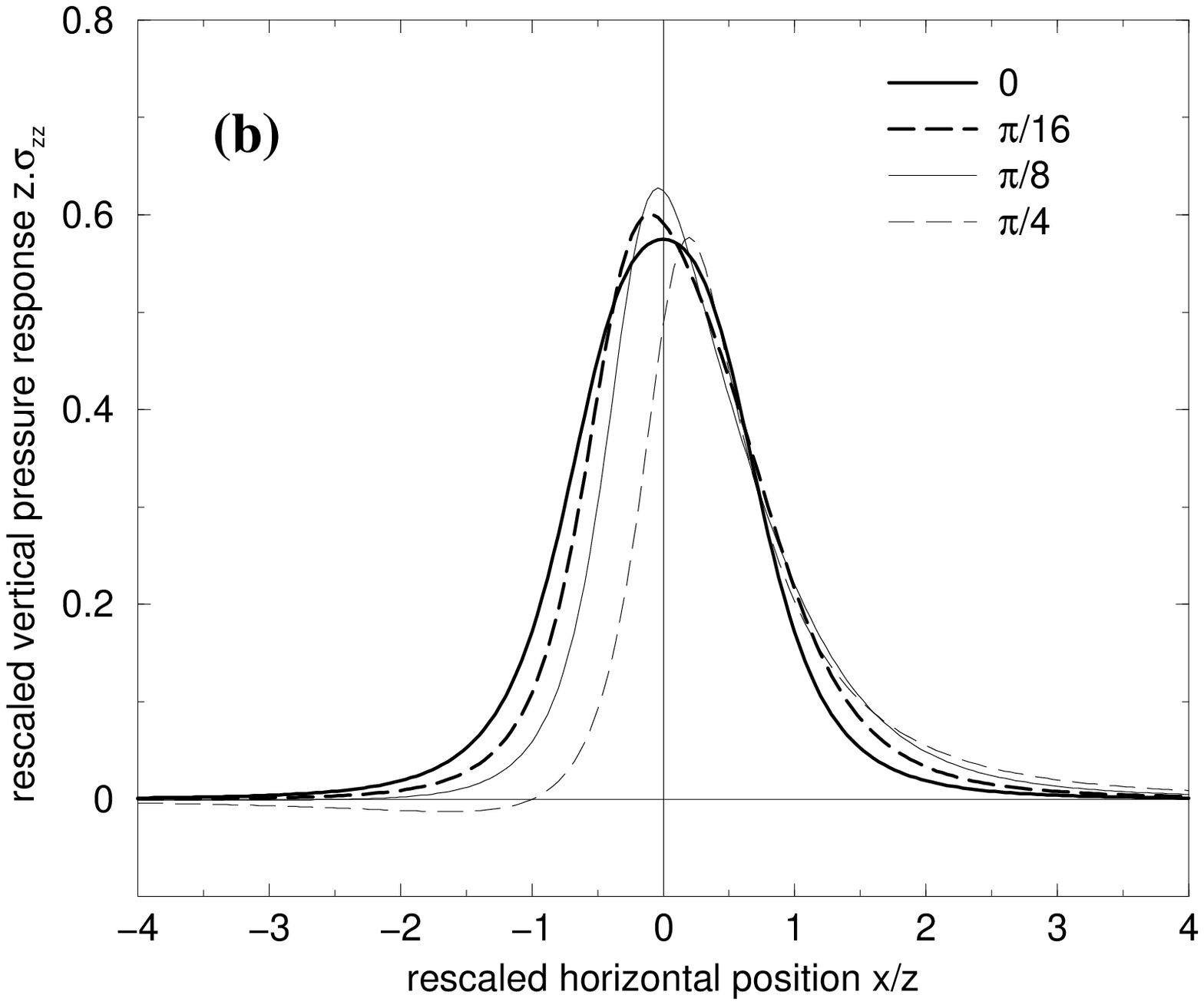} 
\caption{Same graphs as in figure \protect\ref{t0=0caseI}, but this time with 
$\theta_0=\tau$ as indicated in legends. 
\label{t0=taucaseI}} 
\ec 
\efig 

\subsubsection*{Region II}

\bfig[t!]
\bc
\epsfxsize=0.49\linewidth
\epsfbox{szz.t=2.r=1.5.t0=0.eps2}
\hfill
\epsfxsize=0.49\linewidth
\epsfbox{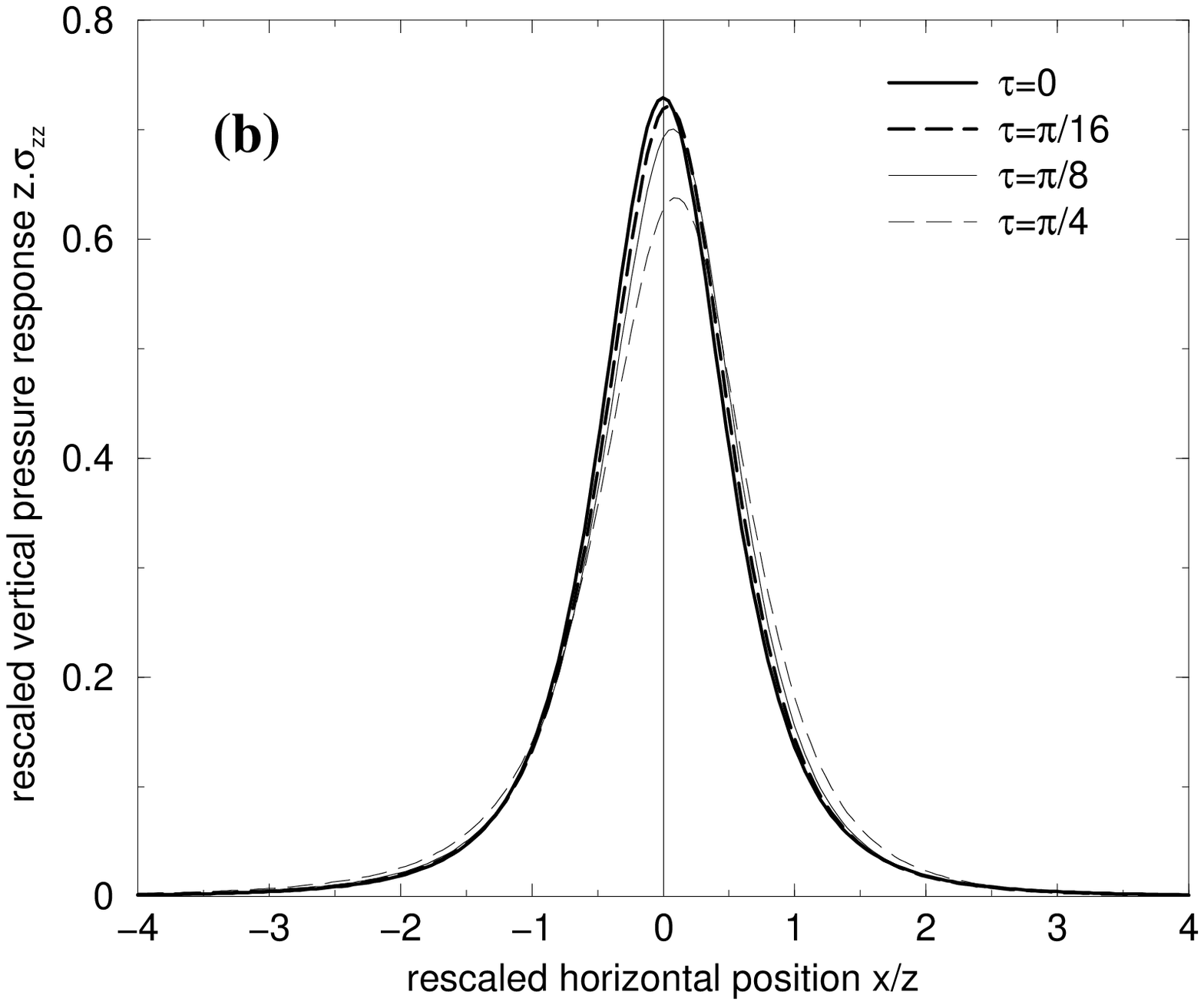}
\caption{{\it Region II}. Response profiles for different values of the anisotropy angle
$\tau$, but with a fixed value for the orientation of the applied force:
$\theta_0=0$. The graph (a) is now for $t=2$ and $r=1.5$, while (b) has been
obtained for $t=0.6$ and $r=0.8$. This time, the response peak can be moved
to the right or to the left with positive values of $\tau$.
\label{t0=0caseII}}
\ec
\efig
\bfig[t!]
\bc
\epsfxsize=0.49\linewidth
\epsfbox{szz.t=2.r=1.5.t0=tau.eps2}
\hfill
\epsfxsize=0.49\linewidth
\epsfbox{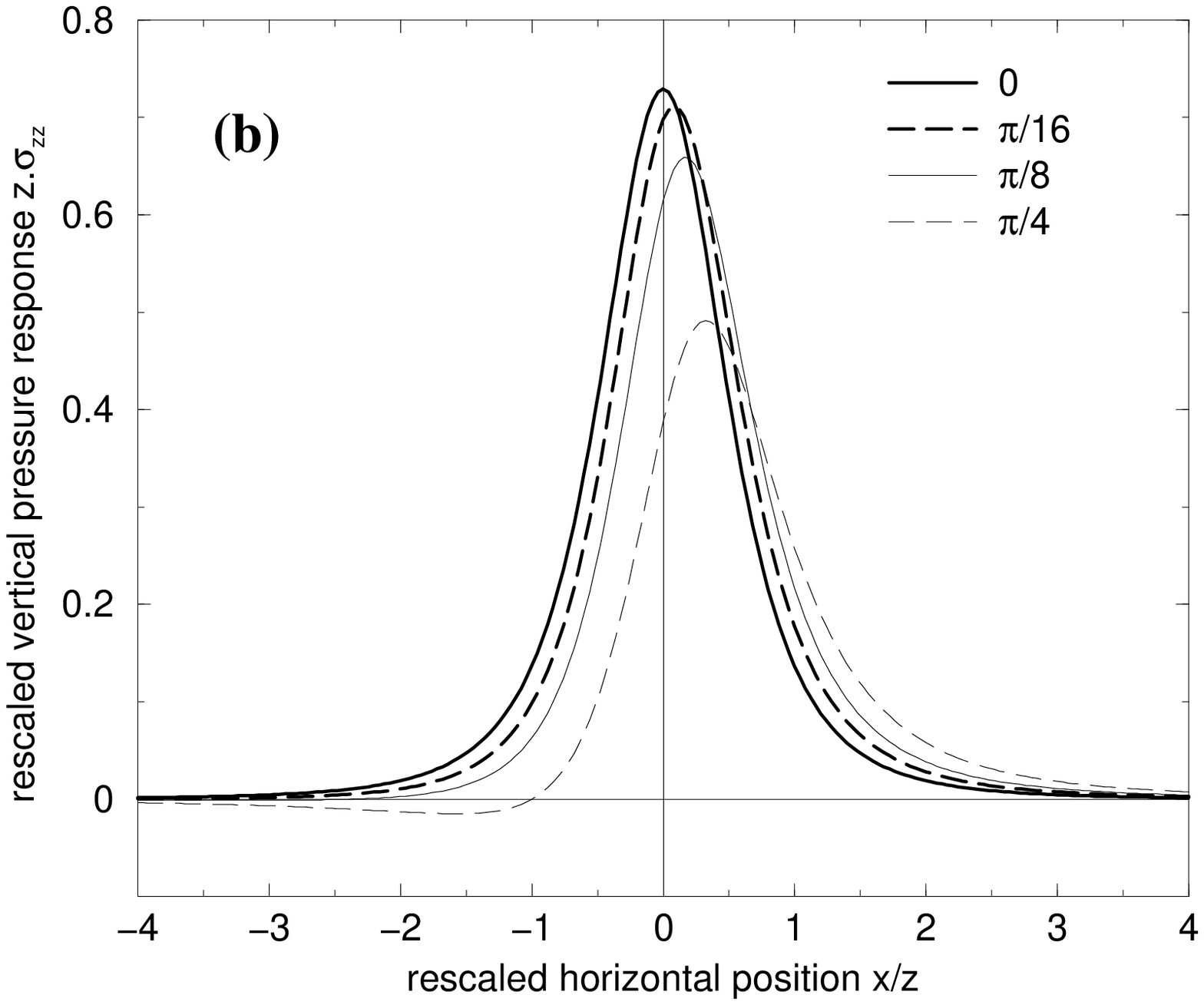}
\caption{Same graphs as in figure \protect\ref{t0=0caseII}, but with
$\theta_0=\tau$ as indicated in legends.
\label{t0=taucaseII}}
\ec
\efig

In region II, where $X_1=-X_4=-i\alpha_1$ and $X_2=-X_3=-i\alpha_2$,
the expressions of the corresponding $Y_k$ involve the quantities
\bea
A_1 & = & \frac{\alpha_1 (1+\tan^2\tau)}{1+(\alpha_1\tan\tau)^2}, \\
B_1 & = & \frac{\tan\tau (\alpha_1^2-1)}{1+(\alpha_1\tan\tau)^2}, \\
A_2 & = & \frac{\alpha_2 (1+\tan^2\tau)}{1+(\alpha_2\tan\tau)^2}, \\
B_2 & = & \frac{\tan\tau (\alpha_2^2-1)}{1+(\alpha_2\tan\tau)^2}, 
\eea
the pressure response having the form 
\bea 
\szz & = &
\frac{F_0}{2\pi} \, \frac{2 z^2}{[(x+B_1z)^2 + (A_1z)^2][(x+B_2z)^2 +
  (A_2z)^2]} \, \left \{ x \sz (A_1+A_2) \right .  \\ \nonumber & &
\left . + z \cz [A_1A_2(A_1+A_2)+A_1B_2^2+A_2B_1^2] + [x \cz + z \sz]
  (A_2B_1+A_1B_2) \right \}.  
\eea 
Again, the expressions of $\sxz$ and $\sxx$ are not shown, but can be
deduced as usual from that of $\szz$.

The Figures \ref{t0=0caseII} and \ref{t0=taucaseII} show the response
profile for different values of the parameters. Depending on
these parameters, the response peak can be moved to the right or to the
left with positive values of $\tau$.

Please note that the response function shown figure
\ref{t0=0caseII}-(b) for $\tau=\pi/4$ also agrees qualitatively with
the experimental findings in \cite{geng}. A more detailed analysis of
their results is certainly worthwhile, also in order to
possibly decide whether region I or II behavior applies for a sheared
two-dimensional layer where the angle of the preferred orientation of
force chains coincides with $\tau=\pi/4$.

\section{Triangular spring networks and anisotropic elasticity}
\label{triang}
\subsection{Triangular spring networks}

To illustrate the previous calculations, it may be useful to construct a
ball-and-spring model with a tunable parameter that allows us to obtain
different relative values of $a$, $b$, $c$, and $d$ above. Here we consider
a triangular lattice of balls with springs connecting all nearest-neighbor
pairs. The lattice may be oriented in either of the two ways shown in figure
\ref{figsprings}, and the springs have stiffnesses $k_1$ or $k_2$ as shown
for the two cases. All springs lying along a given direction have the same
stiffness. We take the equilibrium lengths of all springs to be unity. 

\bfig[t!]
\bc
\epsfxsize=7cm
\epsfbox{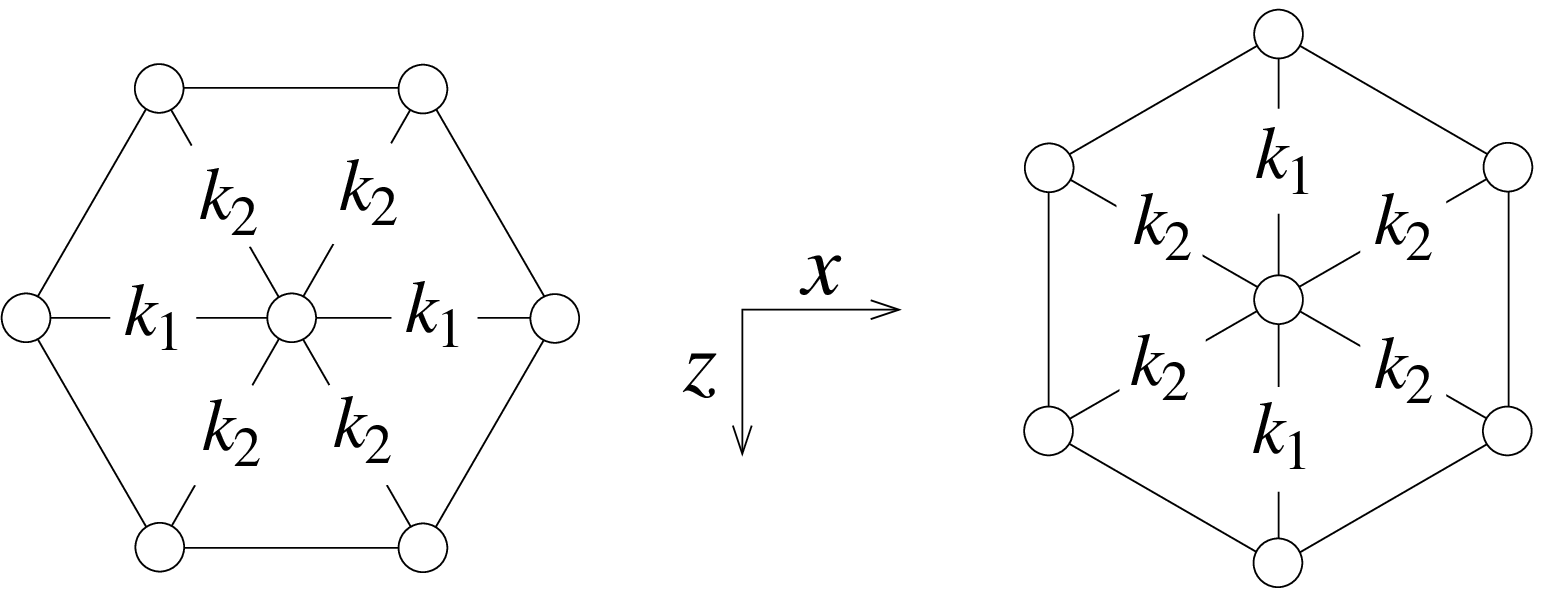} 
\caption{Network of springs of stiffness $k_1$ and $k_2$.
\label{figsprings}}
\ec
\efig

In either orientation, the system has reflection symmetry under
$x\to-x$ and $z \to -z$, but not under rotations; it is described by
an anisotropic stress-strain relation of the form of
${\Lambda}_\dagger$.  We determine the elastic coefficients by writing
down the energy directly for a homogeneous deformation.  Note that the
balls form a Bravais lattice, and hence that their displacements for a
given average strain $u_{ij}$ are simply given by $u_{ij} r_j$, where
${\bf r}$ is the equilibrium position of the ball.  The energy density
can easily be obtained by summing the energies of the three springs
linking the ball at $(0,0)$ to its neighbors along different lattice
directions and dividing by the area of the unit cell, $A=\sqrt{3}/2$.

\subsubsection*{Horizontal orientation of the $k_1$-springs}
For the case where the $k_1$-spring is horizontal,
we find for the energy density:
\begin{equation}
\label{eq:ballspringenergy1}
F = \frac{1}{16 A}
\left[
(8k_1 + k_2) u_{xx}^2 +
9k_2 u_{zz}^2 +
6k_2 u_{xx}u_{zz} +
3k_2 (u_{xz}+u_{zx})^2
\right],
\end{equation}
which corresponds to a matrix ${\Lambda}_\dagger$ with the following coefficients: 
\bea
a & = & \frac{8k_1+k_2}{8A}, \\
b & = & \frac{9k_2}{8A}, \\
c & = & \frac{3k_2}{8A}, \\
d & = & \frac{6k_2}{8A}. 
\eea 
Without loss of generality, we rescale all stiffnesses by a factor $8A/k_2$
and let $k_1/k_2$ be denoted $k$.  The coefficients $r$ and $t$
of equation (\ref{equaX}) are then given by 
\bea
t & = & \frac{1+8k}{9}, \\
r & = & \frac{4k-1}{3}, 
\eea 
which gives $r^2-t=\frac{16}{9}k(k-1)$. We may eliminate
$k$ from these two equations to obtain a trajectory in $(r,t)$ space:
\be
t=\frac{2r+1}{3},
\ee
shown as the plain line in figure \ref{fig0}.

Thus, $k<1$ (weak horizontal springs) corresponds to region I above with (see
equation (\ref{X1and4caseI})):
\bea
\alpha^2 & = & \frac{1}{6} \left ( 4k-1 + \sqrt{8k+1} \right ), \\
\beta^2  & = & \frac{1}{6} \left ( 1-4k + \sqrt{8k+1} \right ).
\eea
As mentioned above, the condition for a double-peaked $\szz$ profile
is $r<0$.  Hence the single-peaked shape of $\szz(x)$ becomes
double-peaked when $k<1/4$, i.e. when the
horizontal springs are substantially softer than the others.

For $k>1$, on the other hand, we are in region II with (see
equation (\ref{X1and4caseII})):
\bea
\alpha_1^2 & = & \frac{1}{3} \left ( 4k-1 + 4\sqrt{k(k-1)} \right ), \\
\alpha_2^2 & = & \frac{1}{3} \left ( 4k-1 - 4\sqrt{k(k-1)} \right ).
\eea
The $\szz$ profile is always a single peaked when the horizontal springs
are stiffer than the others.

\subsubsection*{Vertical orientation of the $k_1$-springs}
For the case where the $k_1$-spring is vertical, we get a matrix $\Lambda_\dagger$
where the coefficients $a$ and $b$ have been swapped from the horizontal case,
i.e. with the following coefficients:
\bea
a & = & \frac{9k_2}{8A}, \\
b & = & \frac{8k_1+k_2}{8A}, \\
c & = & \frac{3k_2}{8A}, \\
d & = & \frac{6k_2}{8A}.
\eea
Again, we rescale the stiffnesses and let $k=k_1/k_2$, 
this time finding
\bea
t & = & \frac{9}{1+8k}, \\
r & = & \frac{3(4k-1)}{1+8k},
\eea
which gives $r^2-t=\frac{144\,k(k-1)}{(1+8k)^2}$. As before, $k$ may be
eliminated to obtain the trajectory in $(r,t)$ space:
\beq
t=-2r+3,
\eeq
now corresponding to the dotted line in figure \ref{fig0}.

For $k<1$, we are in region I with
\bea
\alpha^2 & = & \frac{9k}{1+8k}, \\
\beta^2  & = & \frac{3(1-k)}{1+8k}.
\eea
Again, the single peaked shape of the $\szz$ profile
becomes double peaked when $k<1/4$.

For $k>1$, we have $r^2-t>0$ and we are in region II, with
\bea
\alpha_1^2 & = & \frac{3}{1+8k} \left ( 4k-1 + 4\sqrt{k(k-1)} \right ) \\
\alpha_2^2 & = & \frac{3}{1+8k} \left ( 4k-1 - 4\sqrt{k(k-1)} \right )
\eea

\subsubsection*{Three-body (bond-bending) interactions}
For the spring networks discussed above, the Poisson ratios are not
both adjustable simultaneously.  For the horizontal orientation of
$k_1$ springs, $\nu_x=c/a$ is always $1/3$, while for the vertical
orientation $\nu_z=c/b$ is always $1/3$.  In order to have a
ball-and-spring model on a Bravais lattice in which all elastic
parameters can be varied independently, it is necessary to introduce
three-body interactions.  A straightforward way of doing this is to
assume an energy cost for bond angles that differ from $60^{\circ}$.

\begin{figure}[t!] 
\bc 
\epsfxsize=4cm 
\epsfbox{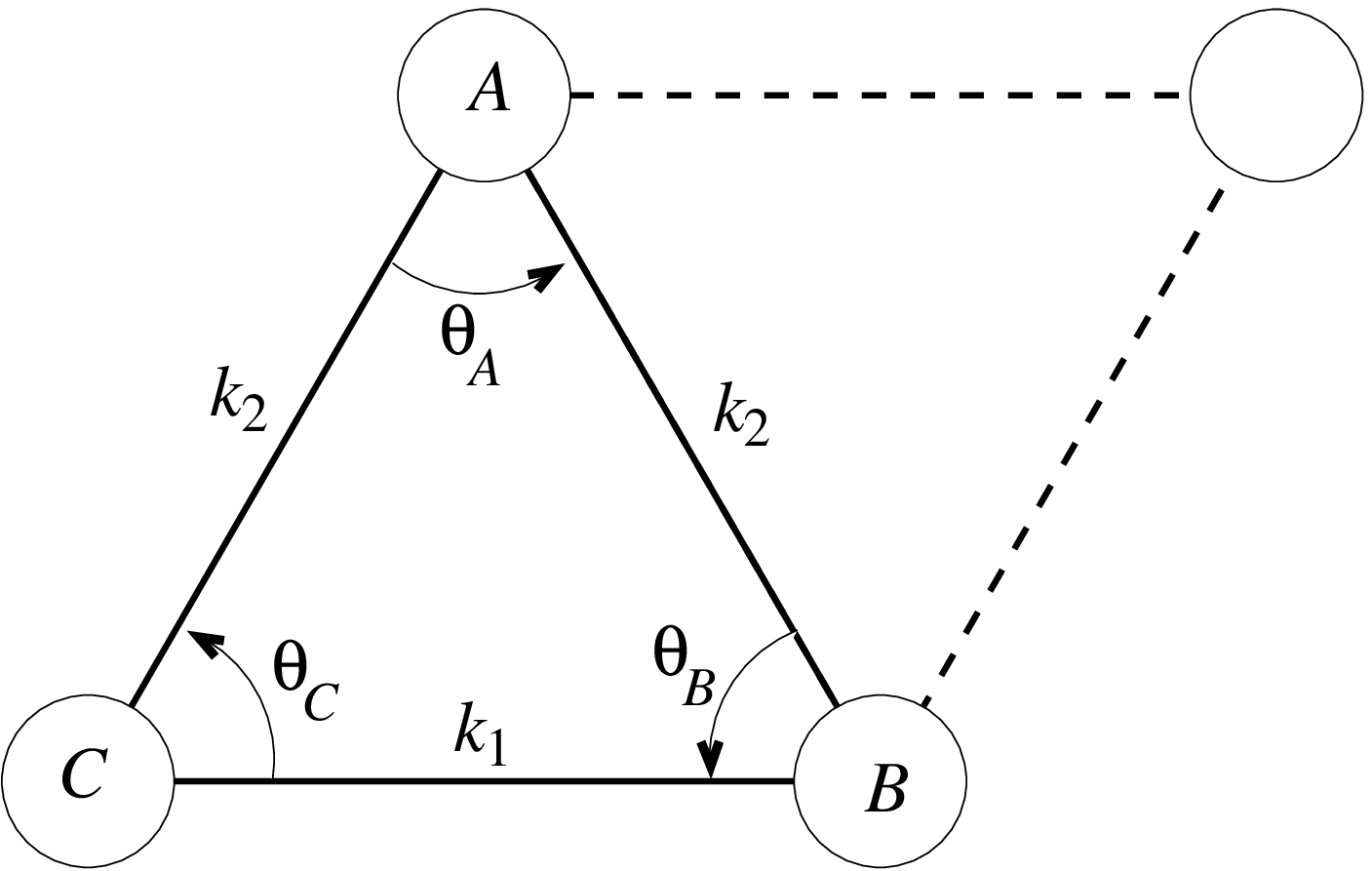} 
\caption{Variables associated with three-body bond-bending interaction. 
\label{figbondbending}} 
\ec  
\end{figure}  

For simplicity, we present an analysis only of the case where the
triangular lattice is oriented so that the $k_1$ springs are
horizontal.  Consider the triangle of balls and springs shown in
figure \ref{figbondbending}.
We define $\theta_Y$ as $\angle XYZ$, measured in the strained
configuration.  For the case of uniaxial symmetry, the energy of the
triangle is determined by two bond-bending stiffnesses $\kappa_1$ and
$\kappa_2$.  For case I we define
\begin{equation} \label{eq:bb}
E_{bb} = (1/2)\left[\kappa_1\left(\theta_A-\frac{\pi}{3}\right)^2 
                   + \kappa_2\left(\theta_B-\frac{\pi}{3}\right)^2 
                   + \kappa_2\left(\theta_C-\frac{\pi}{3}\right)^2\right],
\end{equation}
with $\kappa_1$ assigned to the angle opposite the horizontal edge.
As for equation (\ref{eq:ballspringenergy1}), we take the equilibrium
lengths of the springs to be unity.

Writing expressions for the angles in terms of displacements of the
balls from their equilibrium positions and summing over all triangles,
including the upside-down ones (shown dashed in
figure \ref{figbondbending}) on a homogeneously strained lattice, we
find a contribution to the total energy density of
\begin{equation}
F_{bb} = \frac{3}{8A}\left[
(2\kappa_1+\kappa_2) (u_{xx}^2 + u_{zz}^2)
-2(2\kappa_1 + \kappa_2) u_{xx}u_{zz} +
12\kappa_2 u_{xz}^2
\right]. 
\end{equation}
Adding this contribution to equation (\ref{eq:ballspringenergy1}) gives a
total energy density corresponding to a matrix $\Lambda_{\dagger}$ with
coefficients
\begin{eqnarray}
a & = & \frac{8k_1+k_2+6\kappa}{8A}, \\
b & = & \frac{9k_2+6\kappa}{8A}, \\
c & = & \frac{3k_2-6\kappa}{8A}, \\
d & = & \frac{6(k_2+6\kappa_2)}{8A},  
\end{eqnarray}
where $\kappa\equiv 2\kappa_1+\kappa_2$.  In terms of bulk and shear
moduli and Poisson ratios, we obtain
\begin{eqnarray}
E_z & = & \frac{9k_1 k_2 + 6(k_1+2k_2) \kappa}{(8k_1+k_2+6\kappa)A}, \\
E_x & = & \frac{3k_1 k_2 + 2(k_1+2k_2) \kappa}{(3k_2+2\kappa)A}, \\
G & = & \frac{6(k_2+6\kappa_2)}{8A}, \\
\nu_z & = & \frac{3k_2 - 6\kappa}{8k_1+k_2+6\kappa}, \\
\nu_x & = & \frac{k_2-2\kappa}{3k_2+2\kappa}.
\end{eqnarray}
Note that $E_x \nu_z=E_z \nu_x$, as expected.  Note also that it is
not necessary for $k_1$, $k_2$, $\kappa_1$, and $\kappa_2$ to all be
positive.  Stability (cf. equation (\ref{eq:stability}))
requires only 
\begin{eqnarray}
8k_1+k_2+6\kappa & > & 0, \label{eq:kkstability1}\\
3k_2+2\kappa & > & 0, \label{eq:kkstability2} \\
3 k_1 k_2 + 2\kappa(k_1+2k_2) & > & 0, \label{eq:kkstability3}\\
k_2+ 6\kappa_2 & > & 0. \label{eq:kkstability4}
\end{eqnarray}

>From equation (\ref{r.t}) we find
\begin{eqnarray}
t & = & \frac{8k_1+k_2+6\kappa}{3(3k_2+2\kappa)},  \label{eq:tkk}\\
r & = & 1+\left(\frac{4}{3}\right)
  \frac{3k_1 k_2 + 2\kappa(k_1+2k_2)}{(3k_2+2\kappa)(k_2+6\kappa_2)}
 -\frac{4k_2}{3k_2+2\kappa}.
\end{eqnarray}
By choosing $k_1$, $k_2$, and $\kappa$ we can obtain any positive
value for $t$.  From
Eqs.~(\ref{eq:kkstability1})-(\ref{eq:kkstability4}),  we see that the second
term in the expression for $r$ is positive.  For fixed $t$, we
  can make $r$ arbitrarily large by choosing $\kappa_2$ close to
$-k_2/6$.  The smallest (or largest negative) value of $r$ is obtained
by choosing $3k_1 k_2 + 2\kappa(k_1+2k_2)=0$ (and adjusting
  $k_1$, say, to keep $t$ fixed).  This leads to $r^2-t=0$ and $r<0$,
  demonstrating that the triangular lattice can lie anywhere in region
  I or II.

\subsection{Remarks}

We have seen that classical anisotropic elastic materials
can have double-peaked response functions and that such cases
can be obtained with simple ball-and-spring models.
These calculations explain, for example, the numerical results
of Goldenberg and Goldhirsch \cite{goldhirsch},
without invoking any special considerations on small system sizes.

It is important to note that the response functions for the triangular spring 
networks always lie in the elliptic regime: the peaks broaden linearly with
depth. Thus the observation of a double-peak structure is
{\em not} necessarily an indication of propagative (hyperbolic) response
in an elastic material. However, when the $k_1$ springs are oriented horizontally, and
in the limit where their stiffness tends to zero, the response becomes hyperbolic. 
In this case, one generically expects peaks to broaden diffusively, i.e. like $\sqrt{Dz}$
\cite{pre,Granulab}. Note that in the limit where $k_1 \to 0$, there appears a floppy 
(zero energy) extended deformation mode which, as emphasized by Tkachenko and Witten
\cite{Witten}, naturally leads to a stress-only closure equation and
hyperbolicity. In the phase diagram, figure \ref{fig0}, this limit
corresponds to the point where the straight solid line touches the
boundary curve $t=r^2$.
Note that within this line of thought, one should also expect hyperbolic response 
in elastic percolation networks at the rigidity threshold. In fact, in
the limit $k_1 \to 0$ the triangular network becomes a rhombic
network which is known to become isostatic for a finite system: a single
boundary suffices (say a bottom surface in the slab geometry) in order to suppress the zero mode, and the system becomes rigid \cite{moukarzel}.

\section{Anisotropic Directed Force Chain Networks}
\label{dfcn}
\subsection{Biased scattering}
In \cite{bclo}, a Boltzmann equation for the chain-splitting model was derived for a granular 
medium which is strongly disordered. In the present work, we suppose that the scattering
of force chains by
defects is biased by a preferred orientation of the material, modelled
in terms of a global director ${\bf N}$. We intend to describe systems
possessing a uniaxial symmetry which have undergone compaction or
shearing or which have been constructed by sequential avalanching due
to grains poured from a horizontally moving orifice.

The fundamental quantity is the distribution function $P(f,{\bf n},{\bf r})$, where:
\beq
P(f,{\bf n},{\bf r}) df d{\bf n}d^D r
\eeq
gives the number of force chains with intensity between $f$ and $f+df$, inside the (solid) angle
$d{\bf n}$ around the direction ${\bf n}$,
in a small volume element $d^D r$ centered at ${\bf r}$. Integration of $P(f,{\bf n},{\bf r})$
with respect to $f$ and ${\bf n}$ will yield the density of force chains at the point ${\bf r}$.
\cite{Comment_Socolar}
 The distribution function is defined with respect to an ensemble of different realizations of
 force chains for an assumed uniform
spatial distribution of point defects (of density $\rho_d$), with same boundary conditions.
In the spirit of previous models \cite{bcc,wcc} that give hyperbolic equations for the stresses, a mechanism of
propagation is implemented, but now on the local level of force chains.
 In the analytical model presented here, a pairwise merger of force chain to a single one
 will be neglected. The limitation of this approximation will be discussed below.
Then the distribution function $P(f,{\bf n},{\bf r})$ obeys the following linear equation
\beqa
\label{boltz}
\lefteqn{ P(f_1,{\bf n}_1,{\bf r}+{\bf n}_1 dr)=
\left(1-\frac{dr}{\lambda}\right)P(f_1,{\bf n}_1,{\bf r})} &\nonumber\\
&+&
2\frac{dr}{\lambda}\int \!\!df' \int \!\!df_2\int \!\!d{\bf n}' \int \!\!d{\bf n}_2\,
P(f',{\bf n}',{\bf r})
\Psi({\bf n}'\rightarrow {\bf n}_1,{\bf n}_2|{\bf N})\nonumber\\
&\times&\delta(f_1 \cos \theta_1+f_2 \cos \theta_2- f')
\delta(f_1 \sin \theta_1+f_2 \sin \theta_2)
|\sin(\theta_1-\theta_2)|,
\eeqa
where $\lambda$ is the mean free path of force chains, and is of the
order of $1/(\rho_d l ^{D-1})$ in $D$ dimensions.  The length $l$
represents the average size of a grain.  The equation means the
following: a force chain at some point ${\bf r}+{\bf n}_1 dr$ is
either due to an unscattered force chain, which occurs with the
probability that no scattering occurs times the probability that the
same force chain existed at point ${\bf r}$ (given by the first term
on the r.h.s.  of the equation), or to a scattered force chain.  The
latter occurs with the probability given by the second term on the
r.h.s. of the equation: it is the sum with respect to all intensities
and directions of the incoming (labeled by a prime) and the second
outgoing force chains of the product of the probability for the
incoming force chain to arrive at ${\bf r}$ times the probability of
scattering $\frac{dr}{\lambda}\Psi({\bf n}'\rightarrow {\bf n}_1,{\bf
  n}_2|{\bf N})$.  The delta functions impose conservation of forces,
the factor $2$ accounts for the number of outgoing force chains, and
the factor $|\sin(\theta_1-\theta_2)|$ is convenient to write
explicitly rather than include in $\Psi$.  The dependence of the
scattering probability on ${\bf N}$ requires to consider the outgoing
force chains separately. In the absence of ${\bf N}$ the outgoing
force chains may be treated symmetrically, and one recovers the linear
model for an isotropic medium \cite{bclo}.

The analytical model presented for biased force chain scattering does
not take into account fusion of force chains, which leads in general
to a non linear Boltzmann equation. For an isotropic medium the
consequences of fusion have been discussed for a model where force
chains are restricted to lie on exactly 6 directions \cite{ssc}.  In
this discrete model the validity of the linear approximation was
explicitly shown to be restricted to shallow systems (depths smaller
than a few times $\lambda$) and small forces.  However, preliminary
results on a discrete model with 8 directions suggest that the linear
theory might have a wider scope of application than expected from the
study on the 6-leg model.  More precisely, a proper analysis of the
linear perturbation analysis around the full non-linear solution of
the Boltzmann equation might share, in some regimes, many properties
of the linear solution presented here. In any case, one can see the
present analysis as a shallow layer approximation where the fusion of
chains can indeed be neglected.

Instead of solving equation (\ref{boltz}), we first introduce the scalar
local average force density $F({\bf n},{\bf r})$, i.e. the local scalar
force field per unit volume, defined as
\beq
F({\bf n},{\bf  r})=\int_0^\infty \!\!\!\!\!df\, f P(f,{\bf n},{\bf r}).
\eeq
Then, multiplying  equation (\ref{boltz})
 by $f$, we obtain the following equation for  $F({\bf n}_1,{\bf r})$:
\beqa
\label{force}
\lefteqn{\lambda{\bf n}_1\cdot\nabla_r F({\bf n}_1,{\bf r})=
-F({\bf n}_1,{\bf r})} &\nonumber\\
&+&
2\int \!\!d{\bf n}' \int \!\!d{\bf n}_2 \,
F({\bf n}',{\bf r })
\Psi({\bf n}'\rightarrow {\bf n}_1,{\bf n}_2|{\bf N})\nonumber\\
&\times&\frac{1}{\cos\theta_1-(\sin\theta_1/\sin\theta_2)\cos\theta_2}
.
\eeqa
This equation is identical in form to the
Schwarzschild-Milne equation for radiative transfer \cite{Theo},
  though, unlike the situation in radiative transfer problems, the
  albedo is larger than unity.  Let us note that the possibility to
rewrite the Boltzmann-type equation (\ref{boltz}) in terms of the
force density $F({\bf n},{\bf r})$ is only possible for the linear
model.

>From now on, we take all lengths in units of $l$ which amounts to
formally setting $l=1$.
Now, we introduce physically relevant angular averages
\beqa
p(\vec r) & = & \int \!\! d {\bf n}  \, F(\vec n,\vec r) \\
J_i(\vec r) & = &  \int \!\! d{\bf n} \,  n_i \, F(\vec n,\vec r) \\
\sigma_{ij}(\vec r) & = & D \! \int \!\! d {\bf n}\,  n_i  n_j \, F(\vec n,\vec r),
\eeqa
where $\int d\Omega$ is a normalized integral over the unit sphere.
The field $p$ is the isostatic pressure, while ${\bf J}$ may be
interpreted as the local directed average force chain intensity per
unit surface. Now, given a local snapshot of a force chain network, one
can usually not tell the direction of each chain. Moreover the average
force vanishes everywhere in the system as a consequence of Newton's
third law. The directions of chains are actually determined by the
boundary conditions, say on the top and bottom of a granular layer, which thereby
determine the field ${\bf J}$ in the bulk. It is the propagation of
force chains starting from the boundaries of the system modeled by
equation (\ref{boltz}) which leads to the orientation of the force chain
network.  Finally, the tensor $\sigma$ is the stress tensor.
\subsection{Stress equilibrium at large length scales  }
We now proceed to obtain the equations governing the physically relevant fields introduced above, by 
calculating the zeroth, first, and second moment with respect to $n_i$ of equation (\ref{force}). The equations read as
\beq
\lambda \, \nabla \! \cdot {\bf J} = (c_1-1) p
+c_2 \, \sigma_{NN},
\label{first}
\eeq
\beq
\partial_j \sigma_{ij} = 0,
\label{second}
\eeq
\beqa
\label{third}
\lefteqn{\frac{\lambda}{(D+2)}\left(\delta_{ij} \nabla\cdot{\bf J}+
    \partial_i J_j +\partial_j J_i
  \right)=B_0\,\sigma_{ij}}& \nonumber\\
&+& \delta_{ij} \left( B_1\,\lambda\nabla\cdot{\bf J}+B_2
  \,\sigma_{NN} \right) +{\bf N}_i{\bf N}_j \left(
  B_3\,\lambda\nabla\cdot{\bf J}+B_4 \sigma_{NN}
\right)\nonumber\\
&+&B_5\left( N_i\sigma_{jk}N_k +N_j\sigma_{ik}N_k \right).
\eeqa 
where $\sigma_{NN}={\bf N}\cdot\sigma\cdot{\bf N}$.  The second
equation (\ref{second}) is readily obtained upon averaging, while the first
and third, equations (\ref{first}) and (\ref{third}), are obtained using
an Chapman-Enskog-type expansion of the local average force density
$F({\bf n},{\bf r})$ in terms of the fields $p$, ${\bf J}$, and
$\sigma$ already given in \cite{bclo}: 
\beq \label{CE}  F({\bf n},
{\bf r})= p({\bf r})+ D {\bf n}\cdot {\bf J}({\bf r})+ \frac{D+2}{2}
{\bf n}\cdot\hat{\sigma} ({\bf r})\cdot{\bf n} +\dots 
\eeq
Let us
remark that equation (\ref{second}) gives mechanical equilibrium as expected
and is independent on the specific form of $\Psi({\bf n}'\rightarrow
{\bf n}_1,{\bf n}_2|{\bf N})$.  The validity of the Chapman-Enskog
expansion is based on the assumption that on large enough length scale
an isotropic state is reached. For the case of biased scattering of
force chains considered here, this implies that the bias intensity
must not be too strong. Then the statistical weight of the set of
force chains propagating through the entire system without changing
their direction will not be important.  The limiting case of strong
bias requires a different approach than the one presented here.

The constants $c_\mu$ and $B_\mu$ appearing in Eqs.~(\ref{first})
and~(\ref{third}) respectively are angular integrals involving the
microscopic model for the scattering process, i.e. a specification of
$\Psi({\bf n}'\rightarrow {\bf n}_1,{\bf n}_2|{\bf N})$. A specific
model will be considered in the next section.  If one neglects the
dependence on ${\bf N}$ in the equations above, one recovers the
simpler equations for force chain splitting in an isotropic granular
medium \cite{bclo}.

\subsection{A linear pseudo-elastic theory}
As in the isotropic case, one would like to see if equation (\ref{third})
can be cast into a form where the stress tensor $\sigma_{ij}$ is a
linear function of a pseudo-strain tensor
\beq
u_{ij}\propto\frac{1}{2}\left(\partial_i J_j +\partial_j J_i\right),
\eeq
giving rise to the relation
\beq
\sigma_{ij}=\lambda_{ijkl}u_{kl},
\eeq
where $\lambda_{ijkl}$ is the anisotropic pseudo-elastic modulus tensor.
Similarly to conventional elasticity theory as mentioned in section \ref{ela.general}, 
we will see that the tensor 
$\lambda_{ijkl}$ satisfies
the symmetries given in equation (\ref{sym}). The symmetric form of $u_{ij}$ stems from the symmetries appearing in
the derivation of the large scale equations when carrying out angular
averages, in particular $\int d{\bf n} \, n_i n_j n_k n_l$.
In equation (\ref{third}), the gradients of the field $J_i$ appear only in
combinations such as $\nabla\cdot{\bf J}$ and $\partial_i J_j +\partial_j J_i$.
Please note however that unlike in classical anisotropic linear elasticity theory, in
the present case,
\beq
\label{no.ela}
\lambda_{ijkl}\neq\lambda_{klij},
\eeq
except for certain cases imposed by the details of the scattering process. 
The absence of the symmetry present 
in the classical theory is possible because there is no underlying free energy functional.

The relation between the stress tensor and the pseudo-elastic strain tensor can be derived using the second moment 
equation (\ref{third}).
The latter can be rewritten in the following form
\beq
J_{ij}=B_{ijkl}\sigma_{kl},
\eeq
where 
\beq
J_{ij}=\lambda\nabla\cdot{\bf J}\left[\delta_{ij}\left(\frac{1}{D+2}-B_1\right)-B_3 
N_i N_j\right]+
\frac{\lambda}{D+2}(\partial_i J_j + \partial_j J_i),
\eeq
and 
\beqa
B_{ijkl} &=& 
\frac{B_0}{2}(\delta_{ik}\delta_{jl}
+\delta_{il}\delta_{jk})+B_2 \delta_{ij}N_k N_l,
\nonumber\\
&+& \frac{B_5}{2}(\delta_{jl}N_i N_k + \delta_{jk}N_i N_l
+ \delta_{ik}N_j N_l +\delta_{il} N_j N_k)\nonumber\\
&+& B_4 N_i N_j N_k N_l.
\eeqa
The relation between $J_{ij}$ and $\sigma_{kl}$ can be inverted to give
\beq
\sigma_{ij}=\frac{1}{2}A_{ijkl}J_{kl},
\eeq
where $A_{ijkl}$ has the same form as $B_{ijkl}$ with the constants
$B_\mu$ being replaced by constants $A_\mu$ which are obtained from the relation
\beq
A_{ijkl}B_{klmn}=I_{ijmn}=
\delta_{im}\delta_{jn}+\delta_{in}\delta_{jm}.
\eeq
In particular, one obtains the following relations for the constants $A_\mu$:
\beqa
A_0 &=& \frac{2}{B_0},\\
A_2 &=& -\frac{2B_2}{B_0(B_0+B_2+B_4+2B_5)},\\
A_4 &=& \frac{\left(-2B_4 +\frac{4B_5}{(B_0 + B_5)}(B_2+B_4+B_5)\right)}{B_0(B_0+B_2+B_4+2B_5)},\\
A_5 &=& -\frac{2B_5}{B_0(B_0 + B_5)}.
\eeqa
Now, one can finally determine the pseudo-elastic modulus tensor 
in terms of the tensor $A_{ijkl}$:
\begin{equation}
\label{lambda}
  \lambda_{ijkl} = \frac{\lambda}{D+2}\left( A_{ijkl} + 
                                   \frac{1}{2}A_{ijmm}\delta_{kl}\right)
 - \frac{\lambda}{2}\left(B_1 A_{ijmm}+B_3 A_{ijmn} N_m N_n\right) \delta_{kl}.
\end{equation}
Thus, the pseudo-elastic modulus tensor $\lambda_{ijkl}$ becomes -- via the tensor $A_{ijkl}$ and the constants 
$A_\mu$ -- a function of the constants $B_\mu$ which depend on the specific scattering model used.

In the next section,
a special case will studied which allows us to derive a simple,
but non-trivial equation for the stresses which supplemented by the mechanical equilibrium condition (\ref{second}) 
opens a way to determine the stress tensor, or, put differently, the response function.

\subsection{A microscopic model for force chain splitting in presence of a bias}
\begin{figure}[t!]
\begin{center}
\epsfxsize=8cm
\epsfbox{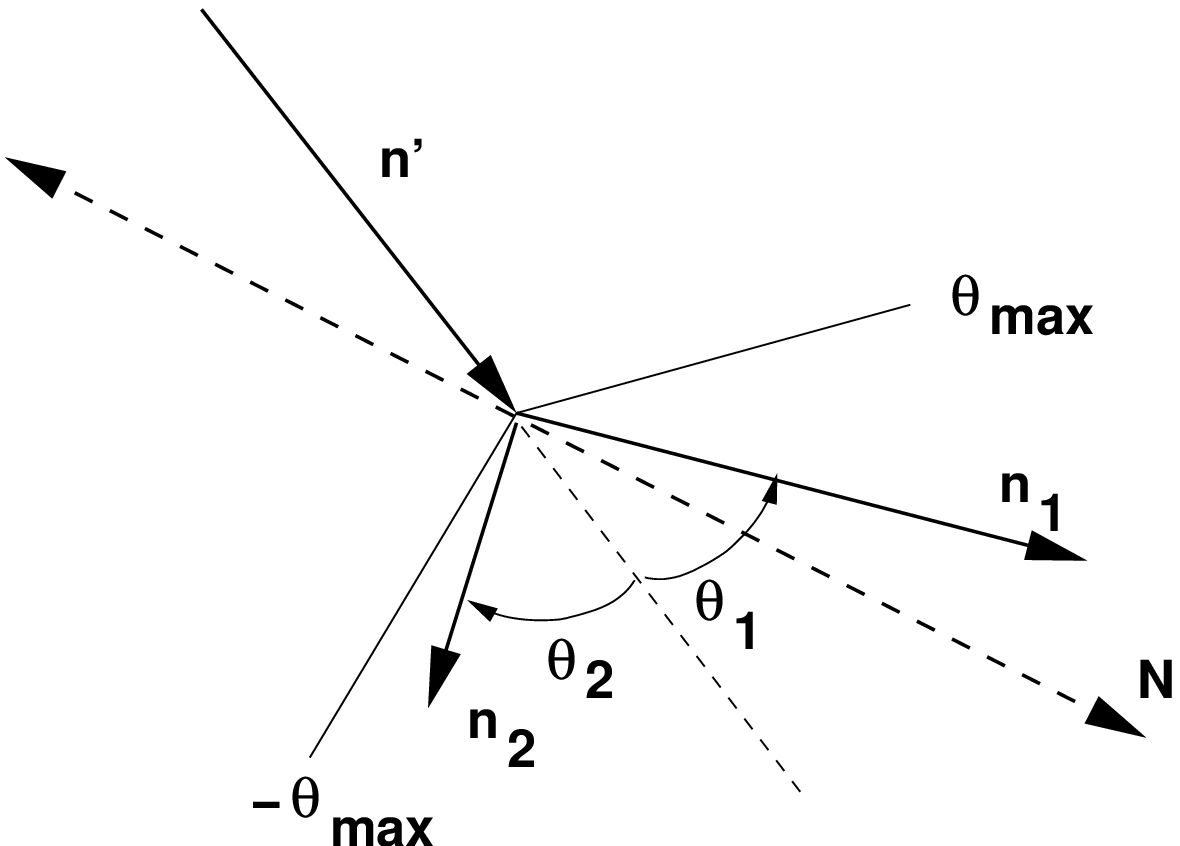}
\caption{The microscopic scattering model. The length of the arrows are different to 
illustrate the amount of force transmitted along the directions.}
\label{rules}
\end{center}
\end{figure}

As mentioned in the previous section, the entries of the pseudo-elastic modulus tensor depend on the 
 specific model for anisotropic scattering which is specified in terms of the scattering cross
section conditional on the 
global director ${\bf N}$,  $\Psi({\bf n}'\rightarrow {\bf n}_1, {\bf n}_2|{\bf N})$.
We have considered a specific model for force chain splitting. It tunes the strength of the bias for scattering 
parallel to ${\bf N}$, using a weight for each outgoing chain proportional to powers of a cosine factor quantifying 
the degree of collinearity with the global director ${\bf N}$ (see figure \ref{rules}).  

For each force chain arriving at a defect in the direction ${\bf n}'$
two outgoing force chains are chosen in the directions ${\bf n}_1$ and
${\bf n}_2$ as follows: the angle of one chain, say number $1$, w.r.t.
the incoming force chain is chosen with weight $\propto ({\bf
  n}_1\cdot{\bf N})^{2p}$, for a positive integer $p$, in the interval
$[0,\theta_{max}]$ (or $[-\theta_{max},0]$) while the other outgoing
chain, say $2$, is chosen uniformly in the interval
$[-\theta_{max},\theta_1]$ (or $[-\theta_1,\theta_{max}]$
respectively). The reason for choosing the direction of the second
chain like this is that the first (biased) chain should carry most of
the intensity of the incoming force.  Increasing $p$ leads to
scattering which is more and more biased in the the direction ${\bf
  N}$.  The form of the scattering cross section is therefore chosen
as \beq
\label{p.noinf}
\Psi({\bf n}'\rightarrow {\bf n}_1, {\bf n}_2|{\bf N})=C_p \left(
\psi(\theta_2|\theta_1)({\bf n}_1\cdot{\bf N})^{2p}
+\psi(\theta_1|\theta_2)({\bf n}_2\cdot{\bf N})^{2p}
\right)
\eeq
The functions $\psi(\theta_i|\theta_j)$ are the respective (uniform) 
probabilities for $\theta_i$ given $\theta_j$ described above.
The constant $C_p$ is a normalization factor which depends on the angle between ${\bf n}'$ and ${\bf N}$ and which is 
determined from
\beq
\label{norm}
\int \!\!d{\bf n}_1 \int \!\!d{\bf n}_2 \Psi({\bf n}'\rightarrow {\bf n}_1, {\bf n}_2|{\bf N})=1,
\eeq
and its explicit form is given in the Appendix A.4.

The simplest choice for the global director 
is ${\bf N}=\hat{z}$, i.e. if force 
chains are scattered preferably downward. 
We might think of a granular layer that has undergone compaction by 
a vertical load. In this case, the matrix $\Lambda_\dagger$ relating the stress and
pseudo-strain tensor has the block-diagonal form as given in equation (\ref{pseudo.ela}).
Any other orientation of ${\bf N}$ can be related to the vertical one
by an appropriate rotation (see section \ref{angle}, equation (\ref{rotation})).

The numerical values of the parameters $r$ and $t$ that determine the shape
of the response function (see figure \ref{fig0}) depend
in the case of the anisotropic linear directed force chain network
model on the constants $B_\mu$ introduced in the previous section. 
The latter are calculated from the above 
microscopic scattering model (see Appendix A) and are listed in
Tab. I-IV of Appendix A for different choices of the maximum angle
$\theta_{max}$ of the scattering cone and different bias intensities $p$.

Interestingly, the roots we find for this scattering model all lie 
in the (elliptic) regions I and II introduced in figure \ref{fig0}. 
Hence, it is possible to find an anisotropic scattering rule that leads
to a two-peak structure of the response function, but in no cases the 
values of $r$ and $t$ have been found to lie in the hyperbolic region. 
Whether this is a limitation of the linear treatment of the {\sc dfcn}, as suggested by the 
analysis of the 6-fold model \cite{ssc}, is at present not settled. Work in
this direction is underway \cite{ustocome}. 

We finish this section with the following
remark. If one identifies the elastic constants of classical 
anisotropic elasticity
theory and their geometrical generalizations obtained for the linear
anisotropic {\sc dfcn}, as we have always done implicitly here, the possible 
range of values which occur for typical granular materials can be discussed.
Experiments indicate that in samples of  sand  which are filled from above and
 where the major principal axis of a stress tensor is in the vertical direction,
 $t=E_x/E_z$ attains values in the range $0.4<t<1$ (see \cite{Garnier}).
 For the  maximum scattering angles plotted in figure \ref{fig0}, the values of $t$
determined from the specific microscopic model for biased force chain scattering used 
here appear to satisfy the  experimental range.
Further information on the construction history of the sand samples, which
affects e.g. the distribution of packing defects or the strength of the
scattering bias, is needed to fully judge the quality of the anisotropic 
{\sc dfcn} model presented here.

\section{Conclusion}

The main objective of this paper was to work out in details the response 
function
to a localized overload in the case of linear anisotropic elastic, or pseudo-elastic 
materials in two dimensions. 

After working out the details of two specific 
microscopic
models, a triangular network of springs and an anisotropic directed force 
network, 
we have shown that the resulting large scale equations can lead to a large 
variety 
of response profiles, summarized in the phase diagram shown in figure 
\ref{fig0} spanned by a two-parameter combination of entries of the
(pseudo-)elastic modulus tensor. 
The one peak structure of conventional (elliptic) isotropic elasticity 
can split into two peaks for sufficiently anisotropic materials. This 
situation occurs as 
soon as the shear modulus $G$ is greater than the ratio $E_x/\nu_x=E_z/\nu_z$ 
of the Young 
modulus and the Poisson ratio (either in vertical or horizontal direction).
This corresponds to an anisotropic material for which vertical stresses are 
easily 
transformed into horizontal strain (large Poisson ratios) and vice versa but 
which strongly resists shear stresses. However, contrarily
to the prediction of `stress-only' hyperbolic models, these two peaks 
generically 
spread proportionally to the height of the layer, and not as the square root 
of the height
for an hyperbolic medium. For the triangular network of springs, there is a 
special 
point, where the lattice loses its rigidity and a soft mode appears, where 
the system
becomes exactly hyperbolic. It would be interesting to exhibit other situations
where these extended soft modes discussed in \cite{Witten} naturally appear; a
possible candidate is a percolating network of springs at rigidity percolation.

For the anisotropic rules of force chain 
scattering that
we have chosen, on the other hand, the directed force network was always 
found to be in the
elliptic regime. This might however be an artifact of the linear 
approximation that we have used and where mergers of force chains are ignored.
Preliminary results suggest that for the full non-linear problem, a genuine 
elliptic to hyperbolic phase transition might take place when the degree of 
anisotropy
is increased, but more work (underway) is needed to confirm this potentially 
interesting result.  

Recent experiments \cite{geng} have not been able so far to
distinguish between a noisy hyperbolic response (where the width of
peaks scales as the square root 
of the height) or anisotropic
(pseudo-)\\elastic response functions. For sheared system where force
chains are preferably oriented at 45 degrees with respect to the
vertical, response functions show a horizontal shift (in the lateral direction 
with respect to the point of applied force) of the maximum, consistent
with the preferred orientation of force chains. We found qualitative
agreement with our findings. More detailed
experiments appear to be necessary to decide on the parameters $r,t$,
i.e. the possible locations in
the phase diagram, figure \ref{fig0}, or put differently on the elastic constants,
corresponding to a particular form
of the response function, if the present (pseudo-)elastic analysis applies.

It would be interesting to extend the present results to three dimensional 
situations in
order to fit the results of experiments on deep sand layers, where a single 
peak response 
function was measured \cite{Reydellet}, and, most importantly, to test the consistency of the effective elastic 
moduli obtained 
from this fit in other geometries (like the sandpile or the silo). It would 
also be 
very interesting to find a way to prepare a disordered granular medium in a 
sufficiently anisotropic 
state such as to observe a two-peak response functions.

\section*{Acknowledgments}

We wish to thank B. Behringer, M. Cates, E. Cl\'ement, C. Gay,
I. Goldhirsch, 
 E. Kolb, D. Levine, J.M. Luck, C. Moukarzel,
 G. Ovarlez, G. Reydellet, D. Schaeffer, R. da Silveira, and
 J.P. Wittmer for very useful discussions.
We thank C. Gay for pointing out the papers by Green et al. to us.
M.O. is very grateful to the Service de Physique de l'\'Etat Condens\'e at
CEA, Saclay, where most of this work was performed, for hospitality and
a stimulating atmosphere and acknowledges financial support by a DFG research fellowship OT 201/1-1. 
J.E.S.S. acknowledges support from NSF through grant DMR-01-37119.

\newpage
\section*{Appendix A: Some integrals for biased linear {\sc dfcn}}
\label{appA}
\subsection*{A.1. Zeroth moment}
First, we propose to calculate the coefficients $c_1$ and $c_2$.
Using the expansion (\ref{CE})
the integral w.r.t. ${\bf n}_1$ of the equation for the force density,
one finds
\beqa
\lambda \, \nabla \! \cdot {\bf J} &=&
-p +2\int d{\bf n}' \int d{\bf n}_1\int d{\bf n}_2
\left[
p+D\,n'_i J_i +\frac{D+2}{2} n'_i \hat{\sigma}_{ij} n'_j
\right]
\nonumber\\
&\times&
\Psi({\bf n}'\rightarrow {\bf n}_1,{\bf n}_2|{\bf N})
\frac{1}{\cos\theta_1-(\sin\theta_1/\sin\theta_2)\cos\theta_2}
\nonumber\\
&=& (k_1-1) p
+k_3 \, \frac{D+2}{2}\hat{\sigma}_{NN},
\eeqa
Please note that a contribution occurs only from terms which are even w.r.t.
${\bf n}'\rightarrow -{\bf n}'$.
The first coefficient is given by
\beq
k_1=
2\int d{\bf n}' \int d{\bf n}_1\int d{\bf n}_2
\Psi({\bf n}'\rightarrow {\bf n}_1,{\bf n}_2|{\bf N})
\frac{1}{\cos\theta_1-(\sin\theta_1/\sin\theta_2)\cos\theta_2}
\eeq
The second coefficient $k_3$ appears when performing a decomposition of the
tensor
\beqa
&&2\int d{\bf n}' \int d{\bf n}_1\int d{\bf n}_2
n'_i n'_j
\Psi({\bf n}'\rightarrow {\bf n}_1,{\bf n}_2|{\bf N})
\frac{1}{\cos\theta_1-(\sin\theta_1/\sin\theta_2)\cos\theta_2}
\nonumber\\
&=&k_3^0 \delta_{ij}+ k_3 N_i N_j
\eeqa
The coefficient $k_3^0$ is irrelevant because
$\delta_{ij}\hat{\sigma}_{ij}=0$ where $\hat{\sigma}_{ij}=\sigma_{ij}-\delta_{ij}p$.
The the coefficient $k_3$ is given by
\beqa
k_3 &=&2\int d{\bf n}' \int d{\bf n}_1\int d{\bf n}_2
\left(
2({\bf n'}\cdot {\bf N})^2-1
\right)
\Psi({\bf n}'\rightarrow {\bf n}_1,{\bf n}_2|{\bf N})
\nonumber\\
&\times&
\frac{1}{\cos\theta_1-(\sin\theta_1/\sin\theta_2)\cos\theta_2}
\eeqa
One finally obtains
\beq
c_1=k_1-\frac{D+2}{2}k_3,\;\;c_2=\frac{D+2}{2}k_3
\eeq
Explicit expressions for the constants $k_1$, $k_3$ are given in
section B.4 which finally will have to be evaluated numerically.

\subsection*{A.2. First moment}
Next, let us derive the equation of mechanical equilibrium (\ref{second}).
Taking the first moment of the force density equation without an external force gives
\beqa
\frac{\lambda}{D}\partial_j \sigma_{ij} &=&
-J_i
+ 2\int d{\bf n}' \int d{\bf n}_1\int d{\bf n}_2
n_{1,i}
F({\bf n}',{\bf r})
\Psi({\bf n}'\rightarrow {\bf n}_1,{\bf n}_2|{\bf N})
\nonumber\\
&\times&
\frac{1}{\cos\theta_1-(\sin\theta_1/\sin\theta_2)\cos\theta_2}
\eeqa
The second term contains the integral
\beq
\int d{\bf n}_1\int d{\bf n}_2
n_{1,i}
\Psi({\bf n}'\rightarrow {\bf n}_1,{\bf n}_2|{\bf N})
\frac{1}{\cos\theta_1-(\sin\theta_1/\sin\theta_2)\cos\theta_2}=an'_i
\eeq
Symmetrizing the integrand w.r.t. to the indices $1$ and $2$ gives
$a=1/2$. This result is independent on the specific form for the
scattering cross section $\Psi({\bf n}'\rightarrow {\bf n}_1,{\bf n}_2|{\bf N})
$. The
remaining integral w.r.t. ${\bf n}'$ yields $J_i$ canceling the first term
$-J_i$ above.

\subsection*{A.3. Second moment}
Finally, we calculate the coefficients $B_\mu$ in the third of the
hydrodynamic equations, equation (\ref{third}).
Let us consider the second moment by multiplying the force density
equation by $n_{1,i}n_{1,j}$ and integrating w.r.t. ${\bf n}_1$.
One obtains the following equation:
\beq
\label{app.2moment}
\lambda D\Gamma_{ijkl}\partial_k J_l =
-\frac{1}{D} \sigma_{ij} + \int d{\bf n}'F({\bf n}',{\bf r})
 I_{ij}({\bf n}',{\bf N})
\eeq
where
\beqa
I_{ij}({\bf n}',{\bf N}) &=& 2 \int d{\bf n}_1\int d{\bf n}_2
n_{1,i} n_{1,j}
\Psi({\bf n}'\rightarrow {\bf n}_1,{\bf n}_2|{\bf N})
\nonumber\\
&\times&
\frac{1}{\cos\theta_1-(\sin\theta_1/\sin\theta_2)\cos\theta_2}
\eeqa
and 
\beq
\Gamma_{ijkl}=\frac{1}{D(D+2)}\left(\delta_{ij}\delta_{kl}+\delta_{ik}\delta_{jl}+\delta_{il}\delta_{jk}\right)
\eeq
The tensor $I_{ij}$ may be decomposed as follows:
\beq
 I_{ij}({\bf n}',{\bf N})=
 K_0 \delta_{ij}
 +K_1 n'_i  n'_j   +
 K_2 \left(
 n'_i N_j + n'_j N_i
 \right)
\eeq
The coefficients $K_0$, $K_1$, and $K_2$ are all functions of
the argument ${\bf n}'\cdot
{\bf N}$ which will be suppressed in the following. As the tensor $I_{ij}({\bf n}',{\bf N})$ should be invariant
w.r.t. to the operation ${\bf N}\rightarrow -{\bf N}$ because the scattering cross section $\Psi({\bf n}'\rightarrow {\bf n}_1,{\bf n}_2|{\bf N})$ is, the functions $K_0$ and $K_1$ are even and $K_2$ is odd under this ``parity'' change.
 They are to be determined by multiplying $I_{ij}$ as follows:
\beqa
I_0=I_{ii}&=&K_0 D+ K_1+ 2K_2 ({\bf n}'\cdot {\bf N})\\
I_1=N_i I_{ij} N_j &=&
 K_0 + K_1({\bf n}'\cdot {\bf N})^2 + 2K_2 ({\bf n}'\cdot {\bf N})\\
I_2=n'_i I_{ij} n'_j &=&
 K_0 + K_1 + 2K_2 ({\bf n}'\cdot {\bf N})\\
\eeqa
The variables $I_0$, $I_1$, and $I_2$ are likewise functions of the argument
${\bf n}'\cdot
{\bf N}$ which is suppressed henceforth.
In the following we consider $D=2$.
The system of equations may then be written in matrix form
\beq
(I_0,I_1,I_2)^T={\bf A}(K_0, K_1, K_2)^T
\eeq
with
\beqa
{\bf A}=
\left(
\begin{array}{ccc}
2 & 1 & 2 \cos\alpha \\
1 & \cos^2\alpha & 2\cos\alpha \\
1 & 1 & 2\cos\alpha
\end{array}
\right)
\eeqa
where $\cos\alpha = ({\bf n}'\cdot {\bf N})$. We eventually want the functions
$K_\mu$ as a function of the integrals $I_\mu$, $\mu=0,1,2$. So we need the inverse matrix
\beqa
{\bf A}^{-1}=
\left(
\begin{array}{ccc}
1 & 0 & -1 \\
0 & 1/\sin^2\alpha & 1/\sin^2\alpha \\
-1/(2\cos\alpha) & 1/(2\cos\alpha\,\sin^2\alpha)
& -\cos(2\alpha)/(2\cos\alpha\,\sin^2\alpha)
\end{array}
\right)
\eeqa
We find
\beqa
K_0 &=& I_0 - I_2\\
K_1 &=& \frac{1}{\sin^2\alpha}(I_2-I_1)\\
K_2 &=& \frac{1}{2\cos\alpha}
\left(-I_0+\frac{1}{\sin^2\alpha}(I_1-\cos(2\alpha)I_2)
\right)
\eeqa
Before writing down the integrals $I_\mu$, let us introduce the vector
\beq
{\bf n}_\perp=\frac{{\bf N}-({\bf n}\cdot{\bf N}){\bf n}}{\sqrt{1-({\bf n}\cdot{\bf N})^2}}
\eeq
Furthermore, let us denote the integrals
\beqa
\left(
\begin{array}{c}
i_0\\
i_1\\
i_2
\end{array}
\right)(\alpha) &=&2 \int d{\bf n}_1\int d{\bf n}_2
\left(
\begin{array}{c}
({\bf n}_1\cdot{\bf n}')^2\\
({\bf n}_1\cdot{\bf n}'_\perp)^2\\
({\bf n}_1\cdot{\bf n}'_\perp)({\bf n}_1\cdot{\bf n}')
\end{array}
\right)
\Psi({\bf n}'\rightarrow {\bf n}_1,{\bf n}_2|{\bf N})\nonumber\\
&\times&
\frac{1}{\cos\theta_1-(\sin\theta_1/\sin\theta_2)\cos\theta_2}
\eeqa
Then, we find for the functions $K_\mu$
\beqa
K_0 &=&i_1\\
K_1 &=&-i_1+i_0-2\;{\rm sgn}(\sin\alpha)\cot\alpha i_2\\
K_2 &=&  {\rm sgn}(\sin\alpha)\frac{i_2}{\sin\alpha}
\eeqa
The transformation from $I_0, I_1, I_2$ to $i_0,i_1,i_2$ is primarily
for technical reasons as in the scattering function the directions
${\bf n}_1$ and ${\bf n}_2$ are parametrized w.r.t. ${\bf n}'$.
We may now proceed to perform the integral on the r.h.s. of equation (\ref{app.2moment})
\beqa
& &\int d{\bf n}' F({\bf n}',{\bf r})
 I_{ij}({\bf n}',{\bf N}) =
\int d{\bf n}'
\left(
p+ DJ_k n'_k +\frac{D+2}{2}\hat{\sigma}_{kl}
n'_k n'_l
\right)
I_{ij}({\bf n}',{\bf N})\nonumber\\ 
&=&
p\left(
\delta_{ij}\int d{\bf n}' K_0
+\int d{\bf n}' K_1 n'_i n'_j
+\int d{\bf n}' K_2 (n'_i N_j +n'_j N_i) 
\right)\nonumber\\
&+&D J_k
\left(
\delta_{ij}\int d{\bf n}' K_0 n'_k
+\int d{\bf n}' K_1 n'_i n'_j n'_k
+\int d{\bf n}' K_2 (n'_i N_j +n'_j N_i) n'_k
\right)\nonumber\\
&+& \frac{D+2}{2}\hat{\sigma}_{kl}
\left(
\delta_{ij}\int d{\bf n}' K_0 n'_k n'_l
+\int d{\bf n}' K_1 n'_i n'_j n'_k n'_l
\right.\nonumber\\
&+&\left.\int d{\bf n}' K_2 (n'_i N_j +n'_j N_i) n'_k n'_l
\right)\nonumber\\
\eeqa
The integrals which are multiplied by $J_k$ give no contribution
because due to their tensorial properties they should all be linear in
${\bf N}$ which means that they are uneven under sign change ${\bf
  N}$. On the other hand, the integrands are even w.r.t to this
operation, which implies that the integrals are zero.

We now further simpflify the integrals w.r.t. ${\bf n}'$ multiplied by $p$ and $\hat{\sigma}_{kl}$ using decomposition according to Cartesian tensors.
The integrals following $p$ are denoted as follows
\beq
\int d{\bf n}' K_0 =\bar{K}_0
\eeq
\beq
\int d{\bf n}' K_1 n'_i n'_j=\bar{K}_{1,a}\delta_{ij}
+\bar{K}_{1,b}N_i N_j
\eeq
\beq
\int d{\bf n}' K_2 (n'_i N_j +n'_j N_i) 
=\bar{K}_{2,a}\delta_{ij}
+\bar{K}_{2,b}N_i N_j
\eeq
The constants are given by
\beqa
\bar{K}_{1,a}&=&\int d\alpha K_1 \sin^2\alpha \\
\bar{K}_{1,b}&=&\int d\alpha K_1\cos(2\alpha) \\
\bar{K}_{2,a}&=& 0\\
\bar{K}_{2,b}&=& 2 \int d\alpha K_2\cos\alpha 
\eeqa
The angular integrations above (and all the ones following below) are understood to be normalized by factors $1/(2\pi)$.
The integrals following $\hat{\sigma}_{\gamma\delta}$ are the following:
\beq
\int d{\bf n}' K_0 n'_i n'_j=
\bar{K}_{0,a}\delta_{ij}
+\bar{K}_{0,b}N_i N_j
\eeq
\beqa
\label{4.n}
M_{ijkl}=\int d{\bf n}' K_1 n'_i n'_j n'_k n'_l &=&
\tilde{K}_1\left(\delta_{ij}\delta_{kl}
+\delta_{ik}\delta_{jl}
+\delta_{il}\delta_{jk}
\right)\nonumber\\
&+&
\tilde{K}_2\left(
N_i N_j\delta_{kl}
+\,perm.\,\dots
\right)\nonumber\\
&+&
\tilde{K}_3\,
N_i N_j N_k N_l
\eeqa
and
\beqa
\label{3.n}
& &\int d{\bf n}' K_2 (n'_i N_j +n'_j N_i) n'_k n'_l
\nonumber\\
&=&
\bar{K}_{2}'\left(
2 N_i N_j\delta_{kl}
+ N_i N_k\delta_{jl}
+ N_j N_k\delta_{il}
+ N_i N_l \delta_{jk}
+ N_j N_l\delta_{ik}\right)\nonumber\\
&+&
2\bar{K}_{2}''N_i N_j N_k N_l 
\eeqa
Let us turn to the first of these three integrals. The coefficients are given by the following integrals
\beqa
\bar{K}_{0,a}&=&\int d\alpha K_0 \sin^2\alpha \\
\bar{K}_{0,b}&=&\int d\alpha K_0\cos(2\alpha) \\
\eeqa
The second integral (\ref{4.n}) giving rise to the coefficients $\tilde{K}_i$
is treated by performing the following contractions:
\beqa
M_1 &=& M_{iijj}=\int d{\bf n}' K_1\\
M_2 &=& M_{iikl}N_k N_l
=\int d\alpha K_1 \cos^2\alpha\\
M_3 &=& M_{ijkl}N_i N_j N_k N_l
=\int d\alpha K_1 \cos^4\alpha\\
\eeqa
In matrix notation, the system of equations we have to invert is the following:
\beq
(M_1, M_2, M_3)^T={\bf B}(\tilde{K}_1, \tilde{K}_2, \tilde{K}_3)^T
\eeq
with
\beqa
{\bf B}=
\left(
\begin{array}{ccc}
D(D+2) & 2D+4 & 1 \\
D+2 & D+5 & 1 \\
3 & 6 & 1
\end{array}
\right)
\eeqa
Then, for $D=2$ one finally obtains for the coefficients
\beqa
\tilde{K}_1 &=& \int d\alpha K_1 \left(
\frac{1}{3}-\frac{2}{3}\cos^2\alpha+\frac{1}{3}\cos^4\alpha
\right)\\
\tilde{K}_2 &=& \int d\alpha K_1 \left(
-\frac{1}{3}+\frac{5}{3}\cos^2\alpha-\frac{4}{3}\cos^4\alpha
\right)\\
\tilde{K}_3 &=& \int d\alpha K_1 \left(
1-8\cos^2\alpha+8\cos^4\alpha
\right)
\eeqa
Finally the coefficients of the third integral (\ref{3.n}) read as
\beqa
\bar{K}_{2}' &=&\int d\alpha K_2 \cos\alpha \sin^2\alpha\\
\bar{K}_{2}'' &=&\int d\alpha K_2 \cos\alpha (1-4\sin^2\alpha)
\eeqa
Next, one collects all coefficients in front of the Cartesian tensors 
on the r.h.s. of the second moment (\ref{app.2moment}).
\beqa
\label{app.2moment.1}
& &\lambda D\Gamma_{ijkl}\nabla_k J_l =
-\frac{1}{D} \sigma_{ij} + \hat{\sigma}_{ij}(D+2)\tilde{K}_1
\nonumber\\
&+&
\delta_{ij}
\left(
a_0 p + a_1\hat{\sigma}_{NN}
\right)
+ N_i N_j
\left(
a_2 p + a_3 \hat{\sigma}_{NN}
\right)\nonumber\\
&+& a_4
(N_j \hat{\sigma}_{ik}N_k
+ N_i \hat{\sigma}_{jk}N_k)
\eeqa
where the coefficients $a_\mu$ are given as follows:
\beqa
a_0 &=& \bar{K}_0 + \bar{K}_{1,a}\\
a_1 &=& \frac{D+2}{2}(\bar{K}_{0,b}+\tilde{K}_2)\\
a_2 &=& \bar{K}_{1,b}+\bar{K}_{2,b}\\
a_3 &=& \frac{D+2}{2}(\tilde{K}_{3}+2\bar{K}''_2)\\
a_4 &=& (D+2)(\tilde{K}_{2}+\bar{K}'_2)\\
\eeqa
When reducing the integrals in terms of the integrals $i_\mu$, one obtains
\beqa
a_0 &=& \int d\alpha \left(i_0 \sin^2\alpha + i_1 \cos^2\alpha -
  i_2\;{\rm sgn}(\sin\alpha) \sin(2\alpha)\right)\\
a_1 &=& \frac{D+2}{2}\left(
\int d\alpha \,
i_2 \;{\rm sgn}(\sin\alpha)\cos(2\alpha) +\tilde{K}_2 
\right)
\\
a_2 &=& \int d\alpha \left(
(i_0-i_1)\cos(2\alpha)+2i_2\;{\rm sgn}(\sin\alpha) \sin(2\alpha)
\right)\\
a_3 &=& \frac{D+2}{2}\int d\alpha \left(
(i_0-i_1)(1-2\sin^2(2\alpha))
\right.\nonumber\\
&+&\left.4i_2\;{\rm sgn}(\sin\alpha) \sin(2\alpha)\cos(2\alpha)
\right)\\
a_4 &=& (D+2)(\tilde{K}_{2}+\frac{1}{2}\int d\alpha\, i_2\;{\rm sgn}(\sin\alpha) \sin(2\alpha) )\\
\eeqa
and
\beq
\tilde{K}_1 =\int d\alpha \left(
(i_0-i_1)\sin\alpha -2 i_2\;{\rm sgn}(\sin\alpha) \cos\alpha
\right)\frac{\sin\alpha}{3}(-1+4\cos^2\alpha)
\eeq
where
\beq
\tilde{K}_2 =\int d\alpha \left(
(i_0-i_1)\sin\alpha -2 i_2\;{\rm sgn}(\sin\alpha) \cos\alpha
\right)\frac{\sin\alpha}{3}(-1+4\cos^2\alpha)
\eeq
Using the equation for the zeroth moment (\ref{first}), we can eliminate $p$ and we obtain the coefficients
$B_0$ through $B_5$
\beqa
B_0 &=& -\frac{1}{D}+(D+2)\tilde{K}_1\\
B_1 &=& \frac{1}{(c_1-1)}\left(a_0-a_1
-(D+2)\tilde{K}_1
\right)\\
B_2 &=& a_1+\frac{c_2}{c_1-1}\left(-a_0+a_1+
(D+2)\tilde{K}_1
\right)\\
B_3 &=& \frac{1}{c_1-1}\left(
a_2 -a_3- 2a_4
\right)\\
B_4 &=& a_3+\frac{c_2}{c_1-1}\left(
-a_2 +a_3 +2a_4
\right)\\
B_5 &=& a_4
\eeqa
Inserting for $c_1$, $c_2$ calculated in section B.1, and for
$\tilde{K}_1$, $\tilde{K}_2$, and $a_0$ through $a_4$ given above, the
coefficients $B_\mu$ are entirely determined in terms of integrals
over the scattering function or in terms of integrals over the
functions $i_\mu$ which have to be evaluated numerically.
Explicit
expressions for the functions $i_\mu$ for a specific scattering model
are given in section B.4. and have been used to yield
the following Tab.s (I-IV) in the main text.

\subsection*{A.4. The scattering model}
We give now the explicit form for the normalization factor $C_p$ of the
microscopic scattering model, equation (\ref{p.noinf}).
Choosing the angle between ${\bf n}'$ and ${\bf N}$ as $\alpha$
one finds the following relation to determine $C_p$:
\beqa
1&=&\int d{\bf n}_1 \int d{\bf n}_2 \Psi({\bf n}'\rightarrow {\bf n}_1, {\bf n}_2|{\bf N})
\nonumber\\
&=&C_p 2 \int_0^{\theta_{max}} \frac{d\theta_1}{\theta_{max}}
\int_{-\theta_{max}}^{-\theta_1}\frac{d\theta_2}{(\theta_{max}-\theta_1)}
\left(
\cos^{2p}(\theta_1-\alpha)+\cos^{2p}(\theta_1+\alpha)
\right)
\eeqa
One finds 
\beq
C_p(\alpha)=\frac{1}{4}\left[\frac{1}{2^{2p}} 
\left(
\begin{array}{c}
2p\\
p
\end{array}
\right)
+\frac{2}{\theta_{max}2^{2p}}
\sum_{k=0}^{p-1}
\left(
\begin{array}{c}
2p\\
k
\end{array}
\right)
\frac{\sin((2p-2k)\theta_{max})\cos((2p-2k)\alpha)}{(2(p-k))}
\right]^{-1}
\eeq
We have mentioned above that all constants of the hydrodynamic equations depend on the parameters
$k_1$, $k_3$, and the integrals of the functions $i_\mu(\alpha)$ for $\mu=0,1,2$. Using the model for the 
scattering cross section introduced in the main text, they read as follows
\beqa
\left(
\begin{array}{c}
k_1\\
k_3
\end{array}
\right) &=&2
\int_{-\pi}^\pi \frac{d\alpha}{(2\pi)}C_p(\alpha)
\left(
\begin{array}{c}
1\\
\cos(2\alpha)
\end{array}
\right)
\int_0^{\theta_{max}} \frac{d\theta_1}{\theta_{max}}
\int_{-\theta_{max}}^{-\theta_1}\frac{d\theta_2}{(\theta_{max}-\theta_1)}
\nonumber\\
&\times&
\frac{\left(
\cos^{2p}(\theta_1-\alpha)+\cos^{2p}(\theta_1+\alpha)
\right)}{(\cos\theta_1\sin\theta_2-\sin\theta_1\cos\theta_2)}
\left(\sin\theta_2-\sin\theta_1\right)
\eeqa
and
\beqa
\left(
\begin{array}{c}
i_0\\
i_1\\
i_2
\end{array}
\right)(\alpha) &=&
2\;C_p(\alpha)
\int_0^{\theta_{max}} \frac{d\theta_1}{\theta_{max}}
\int_{-\theta_{max}}^{-\theta_1}\frac{d\theta_2}{(\theta_{max}-\theta_1)}
\frac{\left(
\cos^{2p}(\theta_1-\alpha)\pm\cos^{2p}(\theta_1+\alpha)
\right)}{(\cos\theta_1\sin\theta_2-\sin\theta_1\cos\theta_2)}
\nonumber\\
&\times&
\left(
\begin{array}{c}
\sin\theta_2\cos^2\theta_1-\sin\theta_1\cos^2\theta_2\\
\sin\theta_1\sin\theta_2(\sin\theta_1-\sin\theta_2)\\
\sin\theta_1\sin\theta_2(\cos\theta_1-\cos\theta_2)
\end{array}
\right)
\eeqa
The choice of signs indicated on the r.h.s. is to be understood as follows. The $+$ sign is used for $i_0$, $i_1$, and the $-$ sign for $i_2$.
Using these expression all constants $c_1$, $c_2$, and $B_0$ through $B_5$ can be determined.

\newpage
\subsection*{A.5. Numerical values of the different coefficients}

\begin{table}[ht!]
\begin{center}
\vspace{0.5cm}
\begin{tabular}{|l|r|r|r|r|r|r|} \hline
\em p & 0 & 1 & 2 & 4 & 6 & 8  \\ \hline
$c_1 -1$ &3.23966 & 5.98386  &6.91022 & 7.66596& 8.00018& 8.19206\\
$c_2$ & $<10^{-9}$ & -2.73543 &-3.6542 & -4.39914& -4.72539& -4.91083\\
$B_0$ & -2.02254&-3.07827 &-3.49472  & -3.87001&-4.04823 & -4.15404\\
$B_1$ & 1.12431& 1.05801 &1.03896  & 1.02225&  1.01434& 1.00966\\
$B_2$ & $<10^{-9}$& 0.478234  & 0.688084& 0.897999& 1.00432& 1.06907\\
$B_3$ & $<10^{-9}$& -0.0871608 &-0.0664648 &-0.0348425 & -0.0166413& -0.00515556\\
$B_4$ & $<10^{-9}$& -0.248513 &-0.968517  & -1.88383& -2.39835& -2.72395\\
$B_5$ & $<10^{-9}$& 1.05321  &1.64421 &2.26475 & 2.58598& 2.7831\\
$a$ & 0.185067&0.250961 & 0.374958 & 0.594681&  0.751558& 0.862493\\
$b$& 0.185067&0.297586 &0.45714& 0.727492& 0.916382&  1.04853\\
$c$&0.432281&0.308721  &0.315767 & 0.368355& 0.416152& 0.452717\\
$c'$&0.432281& 0.971317 &1.48443 & 2.25966& 2.76618 & 3.10848\\
$d$& -0.247214& -0.246906 &-0.270197& -0.311477& -0.341938& -0.364713\\
$r$& 1.0& 0.914026 &0.438156& -0.042155& -0.260547& -0.383077\\
$t$& 1.0& 0.843324 &0.820227 & 0.81744&  0.820136& 0.822574\\
\hline
\end{tabular}
\end{center}
\label{tab1}
\caption{The microscopic constants $c_0$, $c_1$ and $B_\mu$,
  $\mu=0,\dots,5$, the entries $a,b,c,c',d$ of the matrix $\Lambda_\dagger$ calculated
from the microscopic model for scattering for different bias
intensities $p$ where the maximum
scattering angle is $\theta_{max }=\pi/2-0.01$.}
\end{table}

\begin{table}[ht!]
\begin{center}
\vspace{0.5cm}
\begin{tabular}{|l|r|r|r|r|r|r|} \hline
\em p & 0 & 1 & 2 & 4 & 6 & 8  \\ \hline
$c_1 -1$&1.90263&3.3809   &3.90443 &4.36006 &4.57993 & 4.7162\\ 
$c_2$& $<10^{-9}$& -1.4544 &-1.95725 &-2.38459 &-2.58451 & -2.70519\\
$B_0$&-1.34045 & -1.97435& -2.22524 &-2.45813 &-2.57493 & -2.64818\\
$B_1$& 1.20453 & 1.12804 & 1.10869 &1.09218 & 1.08417 &1.0792 \\
$B_2$& $<10^{-9}$&0.302278  &0.418453 & 0.53197  & 0.590517 &0.627352 \\
$B_3$& $<10^{-9}$&-0.0881382 &-0.0775209 &-0.0563067 &-0.0438889 & -0.0353852\\
$B_4$& $<10^{-9}$&-0.158535 &-0.486841 &-0.90403 &-1.14793 &-1.30717 \\
$B_5$& $<10^{-9}$& 0.626319 &0.940457 &1.26503 &1.43651 &1.54521 \\
$a$& 0.339085&0.40072 &0.51712 &0.705856 & 0.828683&0.915888 \\
$b$& 0.339085&0.501589 & 0.681264& 0.952482& 1.1194& 1.23675\\
$c$&0.712094&0.521519  &0.513996 &0.548739 &0.580671 & 0.606106\\
$c'$& 0.712094&1.36669 & 1.89275 &  2.61839&3.04756 & 3.3414\\
$d$& -0.373009& -0.370911 &-0.38917 & -0.419075 &-0.439207 & -0.453324\\
$r$&1.0 &0.868488  &0.57427& 0.252696  &0.0920036 &-0.00397738 \\
$t$&1.0 & 0.7989 &0.75906 &0.74107 &0.740293 &0.740561 \\
\hline
\end{tabular}
\end{center}
\label{tab2}
\caption{The microscopic constants $c_0$, $c_1$ and $B_\mu$,
  $\mu=0,\dots,5$, the entries $a,b,c,c',d$ of the matrix $\Lambda_\dagger$ calculated
from the microscopic model for scattering for different bias
intensities $p$ where the maximum
scattering angle is $\theta_{max }=\pi/2-0.05$.}
\end{table}

\begin{table}[t]
\begin{center}
\vspace{0.5cm}
\begin{tabular}{|l|r|r|r|r|r|r|} \hline
\em p & 0 & 1 & 2 & 4 & 6 & 8  \\ \hline
$c_1 -1$&0.154395 &0.224047 &0.271432 &0.32539  &0.355728  & 0.37555\\
$c_2$&$<10^{-9}$ &-0.0528349  &-0.0821605 &-0.113501 &-0.130596 & -0.14166  \\
$B_0$&-0.199655 &-0.267729 &-0.311707 &0.360256 &-0.386527 &-0.403233 \\
$B_1$&1.79315 &1.70859 &1.69527 &1.68785 & 1.68312&1.67972 \\ 
$B_2$&$<10^{-9}$ &0.0282278 &0.0350513 &0.0396921 &0.0413126 &0.0421027 \\
$B_3$&$<10^{-9}$ &-0.0272506 &-0.0938402 &-0.16046 &-0.192802 &-0.211874 \\
$B_4$&$<10^{-9}$ &-0.0482741 &-0.0321304 &0.000361819 &0.0226003 &0.038231 \\
$B_5$&$<10^{-9}$ & 0.0590451 &0.0753626 &0.0858452 &0.0893064 &0.0908931 \\ 
$a$&5.22475 &4.46981 &3.99399 &3.55095 &3.33471 & 3.20677\\
$b$&5.22475 &5.48863 &5.38669 & 5.23452&5.14104 &5.08701 \\
$c$&7.72907 & 6.02669&5.24234 & 4.56791&4.25716 & 4.07678\\
$c'$&7.72907 & 8.43526& 8.55002& 8.60126& 8.61323& 8.63027\\
$d$&-2.50432 &-2.39596 &-2.11555 & -1.82208  &-1.68225  & -1.60082\\ 
$r$&1.0 &0.682741 &0.765062 &0.912647 &1.00577 & 1.06836\\
$t$&1.0 & 0.814376 &0.741456 & 0.678371& 0.648644& 0.630384\\
\hline
\end{tabular}
\end{center}
\label{tab3}
\caption{The microscopic constants $c_0$, $c_1$ and $B_\mu$,
  $\mu=0,\dots,5$, the entries $a,b,c,c',d$ of the matrix $\Lambda_\dagger$ calculated
from the microscopic model for scattering for different bias
intensities $p$ where the maximum
scattering angle is $\theta_{max }=\pi/4$.}
\end{table}

\begin{table}[t]
\begin{center}
\vspace{0.5cm}
\begin{tabular}{|l|r|r|r|r|r|r|} \hline
\em p & 0 & 1 & 2 & 4 & 6 & 8  \\ \hline
$c_1 -1$&0.0335067 &0.0428648 &0.0524673 & 0.0641093&0.0705982& 0.0747538\\
$c_2$&$<10^{-9}$ &-0.00668779& -0.0121806& -0.018075& -0.020776&-0.0221954\\
$B_0$&-0.0485058 & -0.0601859 &-0.0709311 &-0.0839805&-0.0912928&-0.0959171\\
$B_1$&1.94764 &1.89848& 1.86723&1.86271& 1.86831& 1.87199\\ 
$B_2$&$<10^{-9}$ &0.00497716& 0.00718661&0.00839984& 0.00836975&0.00806044\\
$B_3$&$<10^{-9}$ & 0.0189319  &-0.0293591&-0.104785&-0.149741& -0.177526\\
$B_4$&$<10^{-9}$ &-0.0118706& -0.0131302& -0.00748694& -0.000990488&0.00438443\\
$B_5$&$<10^{-9}$ & 0.010374 &0.0158534& 0.0190278 &0.0189925&0.018225 \\ 
$a$&24.6907& 22.0583 &19.3127&16.6004& 15.1812&14.279\\
$b$&24.6907 & 25.1971&24.085& 22.3923&21.0886&  20.0854 \\
$c$&34.9988 &29.4735 & 25.2402& 21.4431& 19.66&18.5982\\
$c'$&34.9988 &35.9891 & 35.1548 &33.5005& 31.975 &30.7187\\
$d$&-10.308 &-10.0378 &-9.07808 &-7.69791 &-6.91559 &-6.43566  \\ 
$r$&1.0 &0.697308 & 0.677047 &0.784092 & 0.890948 &0.973368\\
$t$&1.0 &0.87543 &0.801857& 0.741342&0.719877&0.710911 \\
\hline
\end{tabular}
\end{center}
\label{tab4}
\caption{The microscopic constants $c_0$, $c_1$ and $B_\mu$,
  $\mu=0,\dots,5$, the entries $a,b,c,c',d$ of the matrix $\Lambda_\dagger$ calculated
from the microscopic model for scattering for different bias
intensities $p$ where the maximum
scattering angle is $\theta_{max }=\pi/8$.}
\end{table}

\clearpage
\newpage

\section*{Appendix B: Response functions}
\label{appB}

\subsection*{B.1. Region I}
The $\sij$ can be expressed as:
\bea
\szz & = & \iqz [ a_3^* e^{-iqx} + a_4 e^{iqx} ] e^{iX_4qz}
       +   \iqz [ a_4^* e^{-iqx} + a_3 e^{iqx} ] e^{iX_3qz}, \\
\sxx & = & \iqz [ (X_3^2 a_3)^* e^{-iqx} + X_4^2 a_4 e^{iqx} ] e^{iX_4qz}
       +   \iqz [ (X_4^2 a_4)^* e^{-iqx} + X_3^2 a_3 e^{iqx} ] e^{iX_3qz}, \nonumber \\
\\
\sxz & = & - \iqz [ (X_3 a_3)^* e^{-iqx} + X_4 a_4 e^{iqx} ] e^{iX_4qz}
       -     \iqz [ (X_4 a_4)^* e^{-iqx} + X_3 a_3 e^{iqx} ] e^{iX_3qz}. \nonumber\\
\eea
The top conditions (\ref{zztop}-\ref{xztop}) allow to calculate the coefficients
$a_3$ and $a_4$. They read:
\bea
a_3 & = & \frac{1}{X_4-X_3} \, \frac{F_0}{2\pi} \, (X_4\cz + \sz), \\
a_4 & = & \frac{1}{X_3-X_4} \, \frac{F_0}{2\pi} \, (X_3\cz + \sz).
\eea
To perform the integrals over $q$, it is useful to define the two following
integrals:
\bea
I_\pm & \equiv & \iqz \cos(qx) \, e^{-\alpha qz \pm i\beta qz}
        =   \frac{\alpha z \mp i\beta z}{(\alpha z \mp i\beta z)^2 + x^2}.\\
J_\pm & \equiv & \iqz \sin(qx) \, e^{-\alpha qz \pm i\beta qz}
        =   \frac{x}{(\alpha z \mp i\beta z)^2 + x^2}.
\eea
We then get
\bea
\szz & = & \frac{F_0}{2\pi} \, \frac{4}{2\beta} \left [
           \beta \cz \frac{I_+ + I_-}{2} +
           \alpha \cz \frac{I_+ - I_-}{2i} +
           \sz \frac{J_+ - J_-}{2i}
           \right ], \\
\sxx & = & \frac{F_0}{2\pi} \, \frac{4}{2\beta} \left [
          \beta (\alpha^2 + \beta^2) \cz \frac{I_+ + I_-}{2} -
           \alpha (\alpha^2 + \beta^2) \cz \frac{I_+ - I_-}{2i}
           \right . \nonumber \\
     &   & \left . -
           (\alpha^2 - \beta^2) \sz \frac{J_+ - J_-}{2i} +
           2\alpha\beta \sz \frac{J_+ + J_-}{2}
           \right ], \\
\sxz & = & \frac{F_0}{2\pi} \, \frac{4}{2\beta} \left [
           (\alpha^2 + \beta^2) \cz \frac{J_+ - J_-}{2i} +
           \beta \sz \frac{I_+ + I_-}{2} -
           \alpha \sz \frac{I_+ - I_-}{2i}
           \right ]. \nonumber\\
\eea

\subsection*{B.2. Region II}
The $\sij$ can be expressed as:
\bea
\szz & = & \iqz [ a_4^* e^{-iqx} + a_4 e^{iqx} ] e^{iX_4qz}
       +   \iqz [ a_3^* e^{-iqx} + a_3 e^{iqx} ] e^{iX_3qz}, \\
\sxx & = & \iqz [ (X_4^2 a_4)^* e^{-iqx} + X_4^2 a_4 e^{iqx} ] e^{iX_4qz}
       +   \iqz [ (X_3^2 a_3)^* e^{-iqx} + X_3^2 a_3 e^{iqx} ] e^{iX_3qz}, \nonumber \\
\\
\sxz & = & - \iqz [ (X_4 a_4)^* e^{-iqx} + X_4 a_4 e^{iqx} ] e^{iX_4qz}
       -     \iqz [ (X_3 a_3)^* e^{-iqx} + X_3 a_3 e^{iqx} ] e^{iX_3qz}. \nonumber\\
\eea
The top conditions (\ref{zztop}-\ref{xztop}) give again
\bea
a_3 & = & \frac{1}{X_4-X_3} \, \frac{F_0}{2\pi} \, (X_4\cz + \sz), \\
a_4 & = & \frac{1}{X_3-X_4} \, \frac{F_0}{2\pi} \, (X_3\cz + \sz).
\eea
In this case, the useful integrals are
\bea
I(\alpha) & \equiv & \iqz \cos(qx) e^{-\alpha qz}
            = \frac{\alpha z}{(\alpha z)^2 + x^2}, \\
J(\alpha) & \equiv & \iqz \sin(qx) e^{-\alpha qz}
            = \frac{x}{(\alpha z)^2 + x^2}.
\eea
We then get
\bea
\szz & = & \frac{F_0}{2\pi} \, \frac{2}{\alpha_2-\alpha_1}
           \left [
           \alpha_2 \cz I(\alpha_1) + \sz J(\alpha_1) -
           \alpha_1 \cz I(\alpha_2) - \sz J(\alpha_2)
           \right ], \nonumber\\
\sxx & = & \frac{F_0}{2\pi} \, \frac{2}{\alpha_2-\alpha_1}
           \left [
          -\alpha_1^2 \alpha_2 \cz I(\alpha_1) - \alpha_1^2 \sz J(\alpha_1))
           \right . \nonumber \\
     &   & \left .
          +\alpha_2^2 \alpha_1 \cz I(\alpha_2) + \alpha_2^2 \sz J(\alpha_2)
           \right ], \\
\sxz & = & \frac{F_0}{2\pi} \, \frac{2}{\alpha_2-\alpha_1}
           \left [
           \alpha_1 \alpha_2 \cz J(\alpha_1) - \alpha_1 \sz I(\alpha_1))
           \right . \nonumber \\
     &   & \left .
          -\alpha_2 \alpha_1 \cz J(\alpha_2) + \alpha_2 \sz I(\alpha_2))
           \right ].
\eea

\end{document}